\documentclass[pdflatex,sn-mathphys-num]{sn-jnl}% Math and Physical Sciences Numbered Reference Style 
\usepackage{graphicx}%
\usepackage[caption=false,font=normalsize,labelfont=sf,textfont=sf]{subfig} %lisäsin
\usepackage{multirow}%
\usepackage{amsmath,amssymb,amsfonts}%
\usepackage{amsthm}%
\usepackage{mathrsfs}%
\usepackage[title]{appendix}%
\usepackage{xcolor}%
\usepackage{textcomp}%
\usepackage{manyfoot}%
\usepackage{booktabs}%
\usepackage{algorithm}%
\usepackage{algorithmicx}%
\usepackage{algpseudocode}%
\usepackage{listings}%
\begin{document}

\title[Defining Quantum Games]{Defining Quantum Games}

%%=============================================================%%
%% GivenName	-> \fnm{Joergen W.}
%% Particle	-> \spfx{van der} -> surname prefix
%% FamilyName	-> \sur{Ploeg}
%% Suffix	-> \sfx{IV}
%% \author*[1,2]{\fnm{Joergen W.} \spfx{van der} \sur{Ploeg} 
%%  \sfx{IV}}\email{iauthor@gmail.com}
%%=============================================================%%

\author*[1,2]{\fnm{Laura} \sur{Piispanen}}\email{laura.piispanen@aalto.fi}
\author[3]{\fnm{Marcel} \sur{Pfaffhauser}}\email{faf@zurich.ibm.com}
\author[3]{\fnm{James} \sur{Wootton}}
%\equalcont{These authors contributed equally to this work.}
\author[4]{\fnm{Julian} \sur{Togelius}}\email{julian.togelius@gmail.com}
\author[5]{\fnm{Annakaisa} \sur{Kultima}}\email{annakaisa.kultima@aalto.fi}
\affil*[1]{\orgdiv{Department of Computer Science}, \orgname{Aalto University School of Science}, \city{Espoo}, \country{Finland}}
\affil[2]{\orgdiv{Department of Applied Physics}, \orgname{Aalto University School of Science}, \city{Espoo}, \country{Finland}}

\affil[3]{\orgdiv{IBM Quantum, IBM Research}, \orgaddress{\street{S\"aumerstrasse 4}, \city{R\"uschlikon}, \postcode{8803}, \state{Zurich}, \country{Switzerland}}}
\affil[4]{\orgdiv{Department of Computer Science and Engineering}, \orgname{New York University Tandon School of Engineering}, \orgaddress{\state{NY}, \country{USA}}}
\affil[5]{\orgdiv{Department of Media}, \orgname{Aalto University School of Arts, Design and Architecture}, \orgaddress{\city{Espoo}, \country{Finland}}}
%%==================================%%
%% Sample for unstructured abstract %%
%%==================================%%

\abstract{In this research article, we survey existing quantum physics-related games and, based on this survey, propose a definition for the concept of quantum games. We define a quantum game as any type of rule-based game that either employs the principles of quantum physics or references quantum phenomena or the theory of quantum physics through any of three proposed dimensions: the perceivable dimension of quantum physics, the dimension of quantum technologies, and the dimension of scientific purposes, such as citizen science or education. We also discuss the concept of quantum computer games, which are games on quantum computers, as well as definitions for the concept of science games. Various games explore quantum physics and quantum computing through digital, analogue, and hybrid means, with various incentives driving their development. As interest in games as educational tools for supporting quantum literacy grows, understanding the diverse landscape of quantum games becomes increasingly important. We propose that the three dimensions of quantum games identified in this article be used for designing, analysing, and defining the phenomenon of quantum games.}

%%================================%%
%% Sample for structured abstract %%
%%================================%%

\keywords{Quantum Games, Game Definitions, Game Studies, Science Games, Serious Games, Gaming, Game Technologies, Quantum Art}

%%\pacs[JEL Classification]{D8, H51}

%%\pacs[MSC Classification]{35A01, 65L10, 65L12, 65L20, 65L70}

\maketitle
\section{Introduction}\label{sec1}
Digital games have become one of the most popular pastimes for a wide variety of people, spanning gender, age and nationality. However, the notion of games and play have been important since the earliest days of mankind \cite{rollefson1992}. Games are played on different devices and platforms: on computers, mobile phones, specialised gaming hardware, and as physical pieces, such as board games and card games. Modes of gameplay also vary between social and solitary activities, even within a single game session. The phenomenon of games is vast, a natural part of human history, and has evolved alongside technological advancements. 

Today, the development of new types of technologies based on quantum physical phenomena, \textit{quantum technologies}, and in particular \textit{quantum computers}, is on the rise, and games have already been designed using these technologies. By the end of 2023, over 260 games referencing quantum physics had been created, drawing the attention of both academics and quantum technology enthusiasts \cite{piispanen2023history,piispanen2023qgj}. To date, there are now well over 350 quantum physics-related games \cite{quantumgames}, and yet no suitable definition has emerged to encompass the vast variety of these games prior to the study done for this article. Previous works have used terms such as ‘quantum games’ or ‘quantum computer games’, referring to \textit{computer games}, where “the rules are based on quantum principles and the games use concepts as superposition, entanglement and the collapse of the wave function” \cite{gordon2010}. In this research article, we provide a broader and deeper definition that, in addition, includes games that run on a quantum computer or use quantum software but might still not fit earlier definitions. We also include games that are not played on (classical or quantum) computers or ones where elements other than the rules are influenced by quantum physical concepts. However, to clarify, this article does not discuss the quantum extension to the mathematical topic of game theory known as \textit{quantum game theory} \cite{eisert1999, perez2024game}. 
 
The ludosphere of quantum physics-related games, meaning the realm of games, design, players, and culture surrounding them \cite{stenros2018}, can only be expected to grow. Attempts to describe the creative process behind quantum game development and therefore initialise an important discussion have so far been limited \cite{archer2022,piispanen2023projects,piispanen2024thesis}, but they underline the need for a vocabulary to discuss these games, their use and development. This article addresses this shortcoming by proposing a vocabulary for dissecting the aspects of quantum physics in games. We examine the history of digital games and quantum physics-related games through the motivations for developing them and offer a closer look at \textit{how} these games reference quantum physics. We explore games inspired by quantum physics, games that teach the theory of quantum mechanics, games that serve the study of quantum physics, and games that use quantum technologies in their implementation. To analyse, discuss, and develop these games meaningfully in the future of quantum game development, we propose a definition of \textit{quantum games} based on three aspects, which we call \textit{the dimensions of quantum games}: 1) the perceivable dimension of quantum physics, 2) the dimension of quantum technologies, and 3) the dimension of scientific purposes (in the field of quantum physical sciences), such as education and citizen science.

\section{Digital Games and Game Definitions}
\label{s:digi}
\begin{figure*}[ht]
\center 
\subfloat[]{\includegraphics[width=0.30\linewidth]{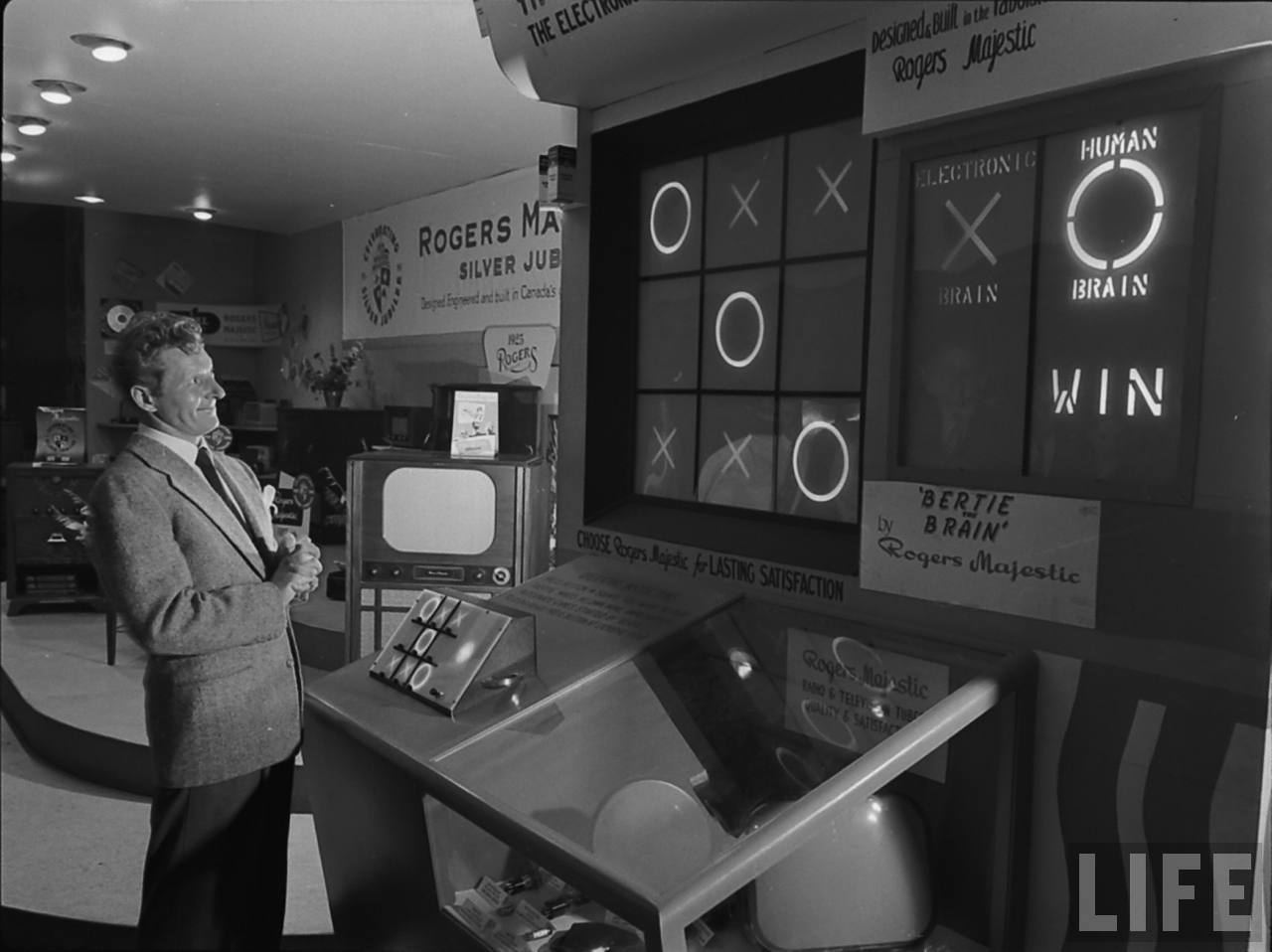}}\,
\subfloat[]{\includegraphics[width=0.23\linewidth]{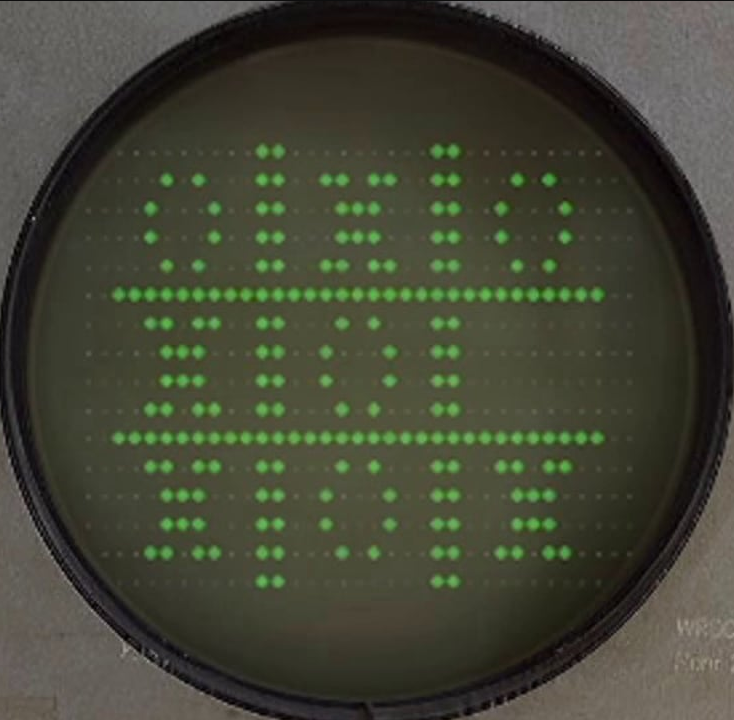}}\,
\subfloat[]{\includegraphics[width=0.336\linewidth]{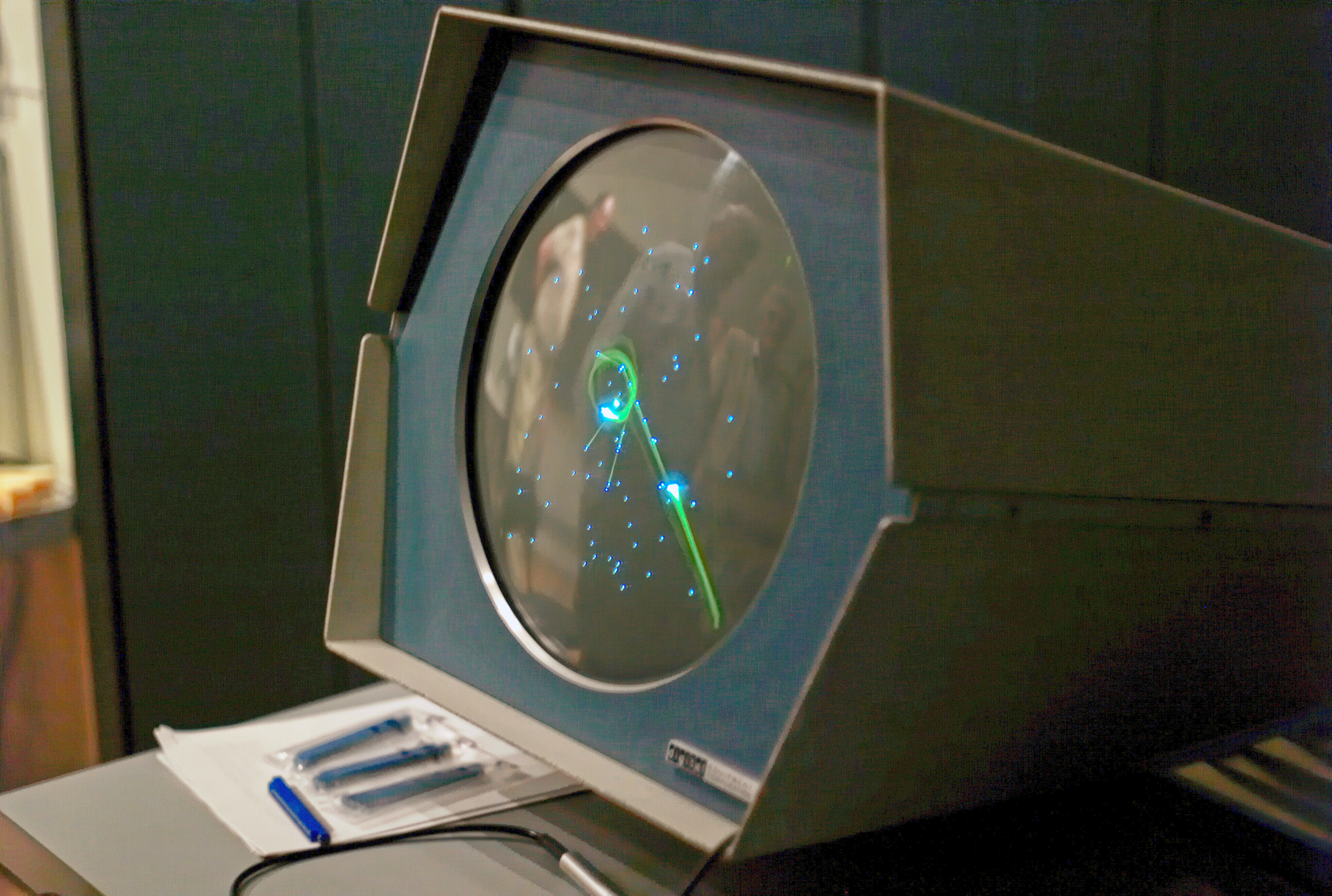}}
\caption{(a) \textit{Life} magazine photo of comedian Danny Kaye standing in front of Bertie the Brain at the Canadian National Exhibition in 1950 (Bernard Hoffman 1950). (b) The game \textit{OXO} on EDSAC programmed by Alexander Shafto Douglas in 1952 (from the \textit{Gaming-History} website ). (c) \textit{Spacewar!} running on PDP-1 (Joi Ito 2007). Both Bertie the Brain and EDSAC demonstrated a game \textit{Tic-tac-toe} against artificial intelligence and were developed primarily to demonstrate the latest advancements in computer technology. \textit{Spacewar!} was a game developed specifically for PDP-1 and aimed to reach a new level of computer entertainment.}
\label{fig:bertiespacewar}
\end{figure*}

Among the multitude of games, digital games play a critical role in our culture today. The early history of computer games dates as far back as the history of the so-called classical (digital) computers. In 1950, one of the first computer games, \textit{Bertie the Brain}, was exhibited at the Canadian National Exhibition (see Figure \ref{fig:bertiespacewar}a). Its main purpose was to demonstrate the use of vacuum tubes together with light bulbs, and did so by playing \textit{Tic-tac-toe}. In 1951 \textit{Nimrod} played the game \textit{Nim}, again using vacuum tubes and light bulbs. The purpose of developing these early computer games was primarily educational, as they illustrated the then-seemingly strange idea of programming principles and algorithms. Another implementation of \textit{Tic-tac-toe} occurred in 1952 at Cambridge University, where the research project \textit{OXO} studied human-computer interaction, again using vacuum tubes but with the addition of cathode ray displays (see Figure \ref{fig:bertiespacewar}b). The project aimed to demonstrate the capabilities of the EDSAC computer (Electronic Delay Storage Automatic Calculator) \cite{oxo}. 

The 1950s saw the development of more games and game-related projects for computers, such as the AI-driven \textit{Checkers} in \textit{The Samuel Checkers-playing Program} by Arthur Samuel at IBM Research \cite{samuel1959}. These early games were primarily about showcasing new technologies, educating people about them, or researching the capabilities of the technology itself. Until the creation of \textit{Tennis for Two} in 1958, computer games were not created purely for entertainment \cite{wolf2008}. 

The value of digital entertainment began to emerge in 1962 when a game was developed not only to test MIT’s new PDP-1 computer and showcase its capabilities, but also to attract a general audience. Moreover, this game was not an adaptation of an existing game, but a completely new game designed specifically for the computer. The game, called \textit{Spacewar!}, depicted space battles (see Figure \ref{fig:bertiespacewar}c). \textit{Spacewar!} was the first game to showcase what computers could do \textit{for games}, paving the way for the evolution of digital games today. 

\subsection{Defining Games}
Defining games and play is not straightforward, and definitions of games are not immutable. Since the formation of game studies as an academic discipline, there has been considerable attention to the definition of games and play \cite{arseth2001}. The first academic definitions for describing games can be said to have been developed in the late 1930s \cite{huizinga1938}, and the most referenced works from the 1950s and 1960s \cite{wittgenstein1953,caillois1961} have inspired modern game scholars to provide updated accounts on the “game definition game” \cite{suits2005,juul2005,stenros2016}.
Some of the most influential game definitions list features or conditions that constitute a game \cite{huizinga1938, caillois1961, abt1970, crawford1984, costikyan1994, avedon1971, suits2005} and a synthesis of these definitions is presented in the popular game design book \textit{Rules of Play} by Katie Salen and Eric Zimmerman: “A game is a system in which players engage in an artificial conflict, defined by rules, that results in a quantifiable outcome” \cite{salen2003}.

The so-called “classic game model” suggested by Jesper Juul in 2005, which focuses on rule-based games, incorporates the player's experience into the definition: “A game is a rule-based system with a variable and quantifiable outcome, where different outcomes are assigned different values, the player exerts effort in order to influence the outcome, the player feels emotionally attached to the outcome, and the consequences of the activity are negotiable” \cite{juul2005}. We note that comprehensively defining what games are is not a straightforward task; therefore, games need to be defined continuously \cite{arjoranta2019}. The definition of quantum games derived in this article is therefore designed to serve the multitude of game formats as widely as possible without touching the definition of games as a phenomenon. In addition, we hope that our definition will also serve other creations, such as \textit{quantum art} in its various forms that combine aspects of quantum physics with art, without touching the definitions of art itself \cite{dickie1969, young1997,cattan2024,crippa2024quantum}. 

\subsection{Defining Science Games}
As the discipline of game studies has matured, we have also developed more vocabulary for different types of games. There are several academic subcategories of games, including leisure games, educational games, and science games. Many, if not most, games labelled as "science games" or “science-based games” are educational games about natural sciences, often designed especially for kids \cite{beier2012, sbasedgames, anupam2022}. Digital science games offer a greater variety of formats and purposes, such as building motivation for careers in science \cite{beier2012}. Still, no common inclusive definition encompasses other purposes. A categorisation of science games proposed by Rikke Magnussen based on the different goals proposed for the player underlines the variety of these games \cite{magnussen2014}. These categories include training, inquiry, professional simulations, epistemic games, embodied system games, and research collaboration games. An example is \textit{citizen science games}, where \textit{gamification} is used to boost citizen science initiatives.

\textit{Gamification} is the introduction of playful, game-like elements into an originally non-game context, such as scientific research \cite{hamari2019}. Tasks like data processing, which require numerous work hours, can be outsourced to citizen volunteers through \textit{citizen science games}. In practice, this means that these individuals collect, manipulate, and provide valuable data to support university-level research. Well-designed games offer intuitive problem-solving tasks that are directly connected to the model being researched and serve as tools for harnessing human pattern recognition skills. Examples of such projects include citizen science initiatives focused on protein folding, as in \textit{Foldit}; aligning multiple DNA sequences, as in \textit{Phylo}; and other projects such as \textit{EteRNA} and \textit{Galaxy Zoo} \cite{foldit, phylo, eterna, raddick2009, curtis2014}.\\

To find an inclusive definition for science games, we follow the definition for a \textit{science game jam} event: “\textit{a game jam where the collaboration between game makers and scientists is facilitated in order to produce games contributing to scientific work, directly (such as helping to solve research questions) or indirectly (such as building awareness or teaching a scientific topic).}” \cite{kultima2021qgj}. Similarly, we propose that a \textit{science game} could be defined as \textit{a game contributing to scientific work, either directly (such as helping to solve research questions) or indirectly (such as building awareness, training, or teaching a scientific topic).} Games designed to teach scientific topics or provide training would still fall under the definition of science games as a subgenre. However, this definition would in addition incorporate games like citizen science games under the term \textit{science games}. 

Defined as such, science games fall within the broader context of \textit{serious games} or \textit{applied games}. Games titled as \textit{serious games} are defined as those designed for specific purposes beyond mere entertainment \cite{sawyer2007, djaouti2011}. The academic focus has again been on educational games; Serious games have been described as a genre that “explicitly focuses on education” and as “tools for the development of creativity and innovation competences” \cite{ratan2009, rodriguez2018}. Yet, we want to emphasise that this category also includes games focused on social impact, solving research problems, educating the general public, and training workers to perform specific tasks on job, to give a few examples \cite{nagarajan2012}.\\% With these clarifications we may now with confidence say that citizen science games are serious games and science games through the gamification of citizen science.

While defining the phenomenon of games is a challenge on a theoretical level, keeping up with the multitude of types of play is also difficult \cite{stenros2018}. However, conceptual tools such as definitions will help us improve our understanding of the expanded field of games and play and offer valuable perspectives in operationalising our research questions. Therefore, it is important to continue to have the discussion on defining games, especially to offer definitions for new and emerging types of games. Motivated by this, in the following section, we will pave the way to discuss the quantum physics-related elements in games.

\section{Quantum Physics and Games} 
\label{qpgames}
Quantum physics deals with the subatomic scales of the smallest particles—atoms, electrons, protons, and photons — and describes fundamentally probabilistic phenomena that we do not observe in our everyday lives \cite{busch1995, zeilinger1999}. Still, without quantum physical phenomena, we would not have lasers, transistors, superconducting materials, or modern-day innovations like the tiny computers we have in our mobile phones and smartwatches.

The development of \textit{quantum technologies} -- technologies which rely on quantum physical phenomena -- began its theoretical foundation in the early 1900s and is now at the stage of entering large markets with devices that allow for the manipulation of quantum phenomena with unprecedented precision and fault tolerance \cite{dowling2003, deutsch2020, acharya2024}. One such phenomenon is quantum superposition, in which a quantum mechanical object can exhibit multiple states or qualities simultaneously. Another is quantum entanglement, which describes the inherent property of two or more quantum objects acting as a single entity, with their measurement values implicating strong correlations. 

Recent attention has focused specifically on quantum computers, which rely on these and other quantum physical phenomena for their computing power \cite{benioff1980, deutsch1985, divincenzo2000, preskill2023quantum, nofer2023quantum}. The basic units for a quantum computer are quantum bits, or \textit{qubits}. Qubits operate fundamentally differently from classical bits, as their states can be manipulated and controlled to exploit properties like quantum entanglement and quantum superposition in computational processes \cite{nielsen2010}. This is the unique advantage of quantum computing over classical computation.

It is possible to simulate quantum computing classically to an extent, but for quantum computing on a large number of qubits a quantum computer is needed as the required resources of a classical nature grow exponentially with the number of qubits \cite{feynman,zhou2020}. Scalable and reliable quantum computers are promised (or hyped \cite{svozil2016}) to offer a considerable advantage over even the most prominent supercomputers when it comes to specific problems, such as optimising a route with several stops, conducting efficient database searches, or determining structures of proteins \cite{grover1996,marsh2020, robert2021}. Quantum computing holds particular promise in fields such as finance, drug development, industrial optimisation problems, molecular biology, and cryptography \cite{quantumalgorithmzoo, shor1994, outeiral2021, nofer2023quantum}. \\

\begin{figure*}[ht]
\center 
\includegraphics[width=0.30\linewidth]{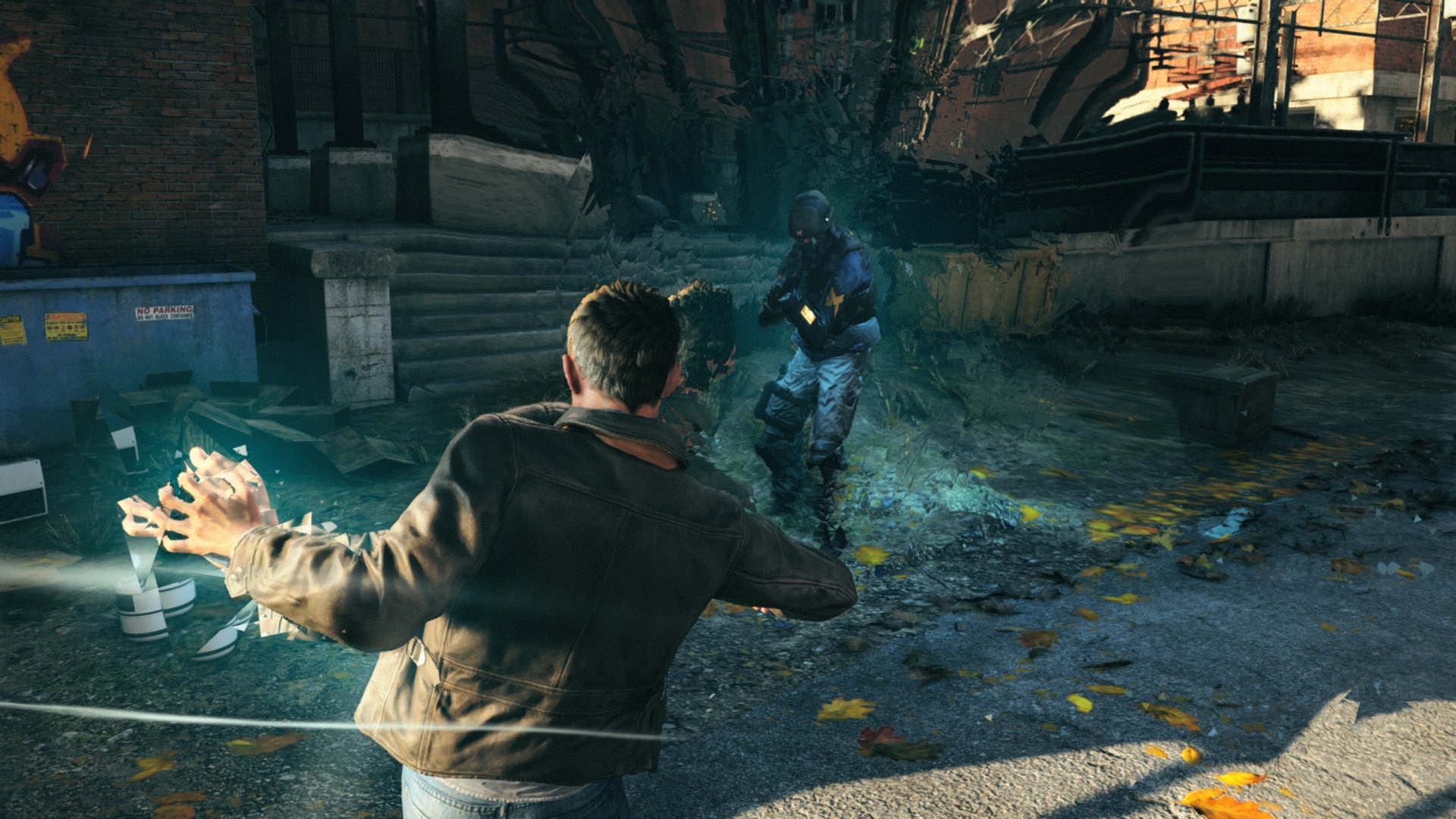}\,
\includegraphics[width=0.30\linewidth]{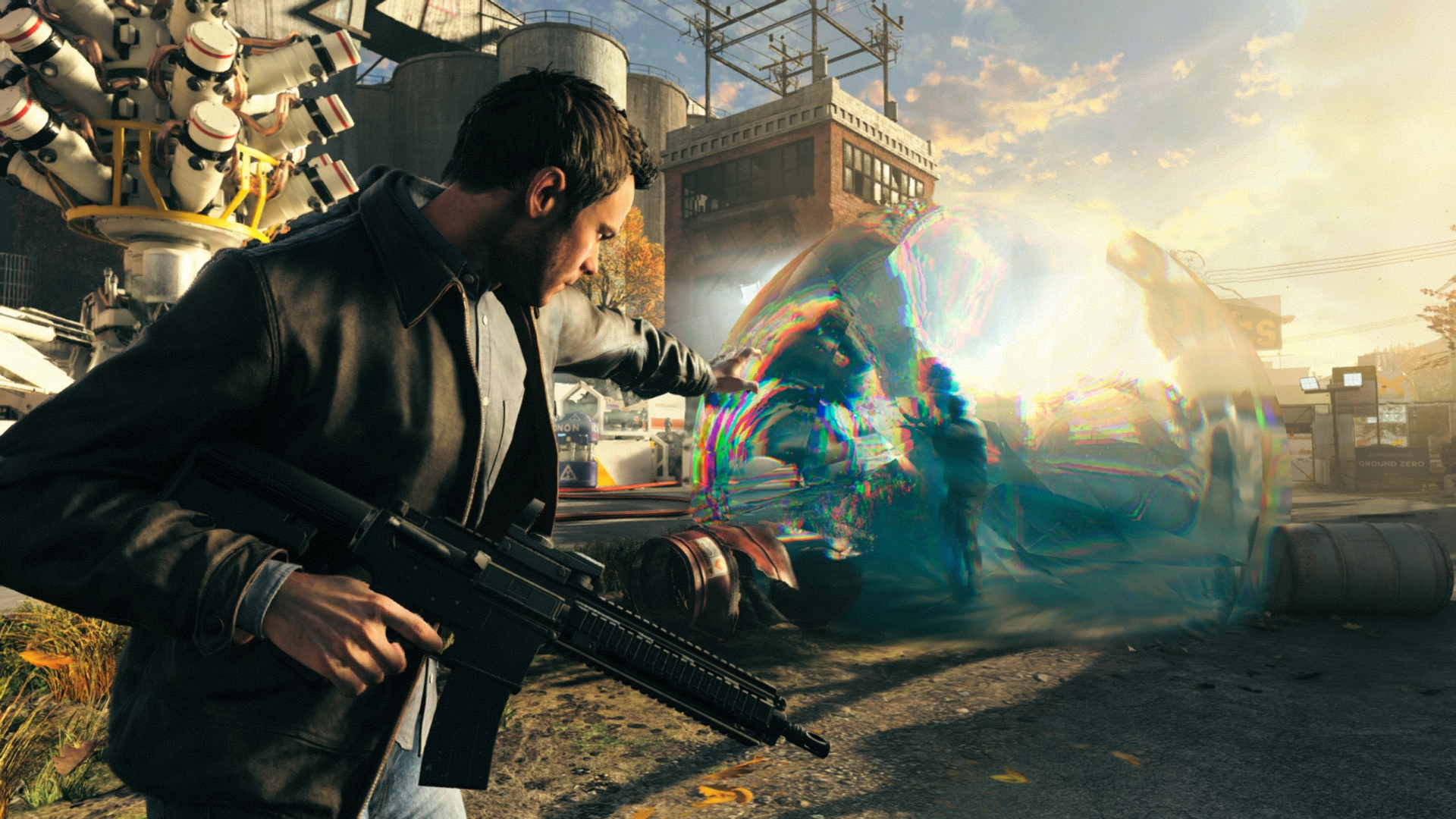}\,
\includegraphics[width=0.30\linewidth]{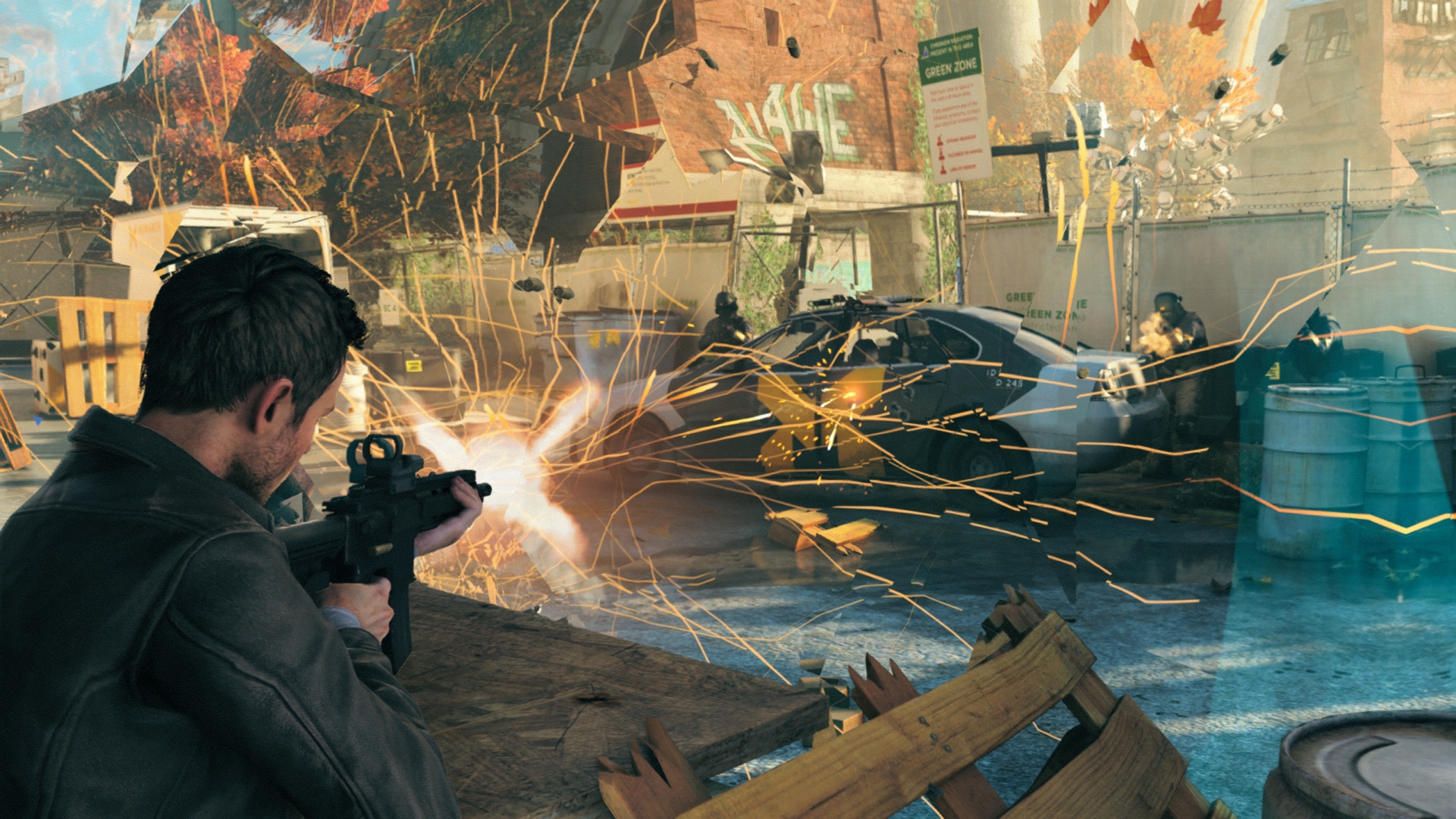}
\caption{Screenshots from game \textit{Quantum Break} (Remedy Entertainment, Press kit (2016)). \textit{Quantum Break} is a science fiction action-adventure third-person shooter where the protagonist is able to control the flow of time.}
\label{screenshotsqb}
\end{figure*}

%\section{Quantum Physics and Games} %%% was here!
One of the earliest, if not the first, commercial game bearing the word “quantum” in its title was \textit{Quantum}, produced for Atari in 1982. In the game, the player enters a subatomic world to capture particles using the optical trackball included in the arcade machine. \textit{Game Over II}, a space adventure-shooter by Dinamic Software released in 1988, had a C64 release titled \textit{Quantum}, but the name was the only connection the game had to quantum physics. Most early games and other forms of entertainment, including movies and series, that referenced quantum physics, have mainly been using the word “quantum” to boost a sci-fi theme. One of the most pressing concerns is that fiction related to quantum physics is often filled with misconceptions, and may thus create fantasy by replacing magic with “quantum” instead of engaging with the theory \cite{svozil2016}. A more recent commercial game referencing quantum physics, \textit{Quantum Break}, serves as something of an exception, as in the development of the game the developers consulted a quantum physicist (see Figure \ref{screenshotsqb}) \cite{rasanen2016,kamen2016}. It must be noted though, that the game is not designed to serve any educational purposes.

\subsection{Simulating Quantum Behaviour}
\label{simul}
Notably, digital games offer greater potential for games than mere inspiration. Numerically simulated quantum principles can be incorporated into game mechanics, allowing players to interact with quantum phenomena rather than encounter misleading fiction \cite{piispanen2024thesis}. Interactive tools that simulate and visualise quantum physical phenomena have taken advantage of this opportunity, successfully strengthening the learning and mental model-building process \cite{kohnle2012, passante2019}. 

This brings us to one of the most popular reasons for building games for quantum physics: education. Quantum mechanics is primarily taught in universities through books, lectures, and now also e-courses. However, since we have no direct reference to quantum phenomena in our everyday lives, studying these abstract concepts through traditional methods often leads to difficulties or even misconceptions \cite{singh2001, singh2008, zhu2012}, which offers great challenges to the progress of quantum literacy \cite{nita2021}. Games offer captivating ways to introduce new concepts and teach the basics of quantum mechanics, inspiring the development of learning platforms and outreach events that integrate games with learning materials \cite{scienceathome, qplaylearn, woottonHelloQuantum, seskir2022, faletic2023contributions, anttila2024can, gaunkar2024game}. Quantum physics-related games have for some time now garnered significant academic attention, especially for their educational applications \cite{peng2014,cantwell2019quantum,anupam2020,lopez2020encrypt,nita2021,chiofalo2022games,seskir2022,montagnani2023,xenakis2023quantum,escanez2024}, but the idea of using numerical simulations in aiding students in the study of quantum physics is not new. Educational simulation tools have proven to be efficient for educational purposes in teaching quantum physical concepts \cite{muller2002, kohnle2012, keebaugh2019, passante2019,hu2024student} and planning research settings, as shown by the \textit{Science at Home} team with their study on the visualisation and simulation tool \textit{Quantum Composer} \cite{ahmed2021}.\\

\begin{figure*}[ht]
\center
\subfloat[]{\includegraphics[width=0.7\linewidth]{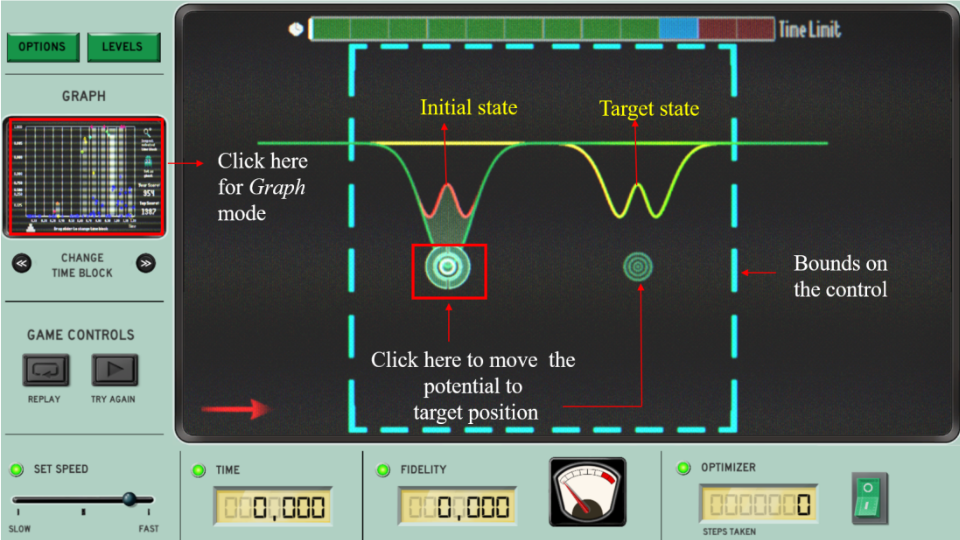}}\\
\subfloat[]{\includegraphics[width=0.45\linewidth]{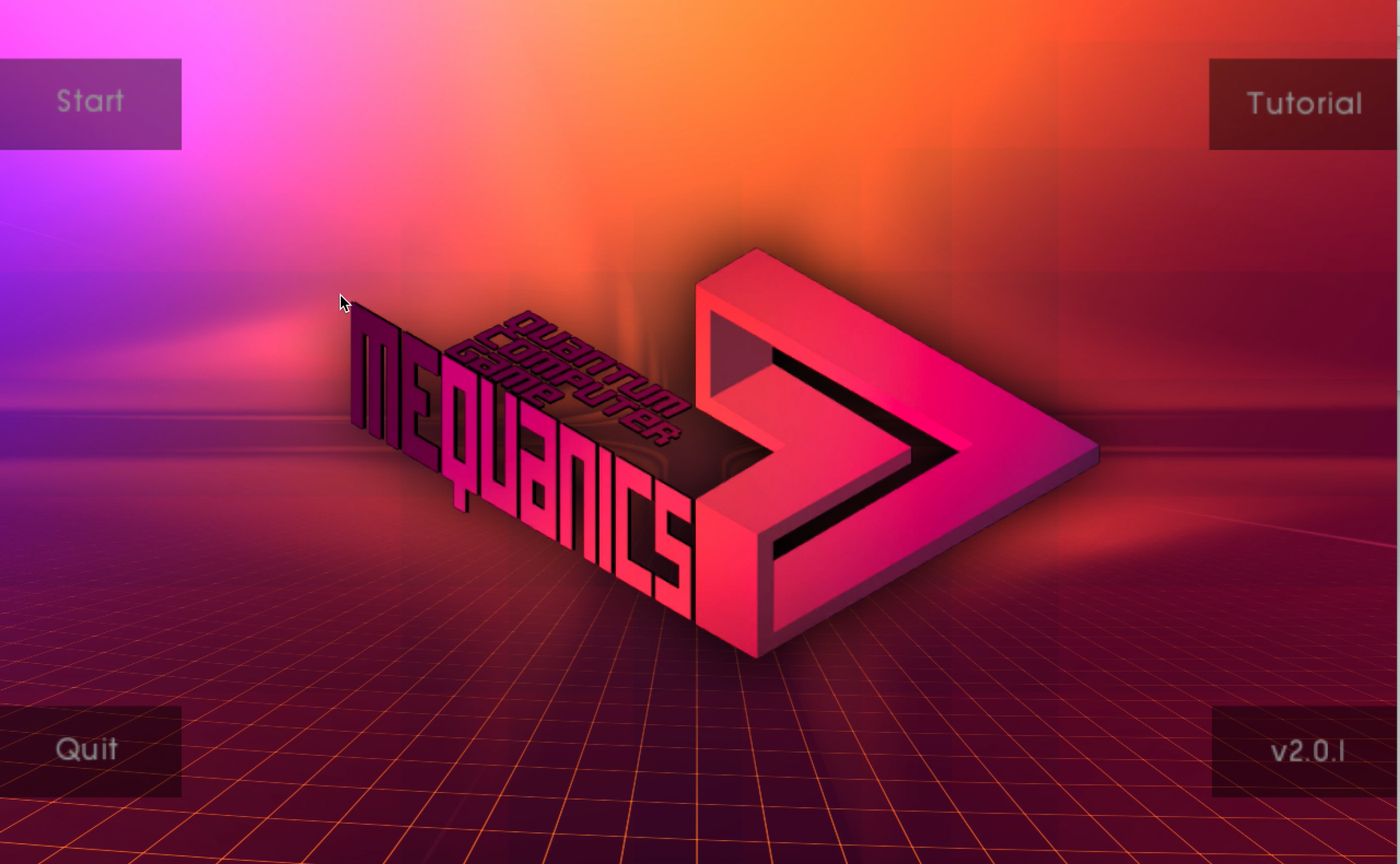}\,
\includegraphics[width=0.45\linewidth]{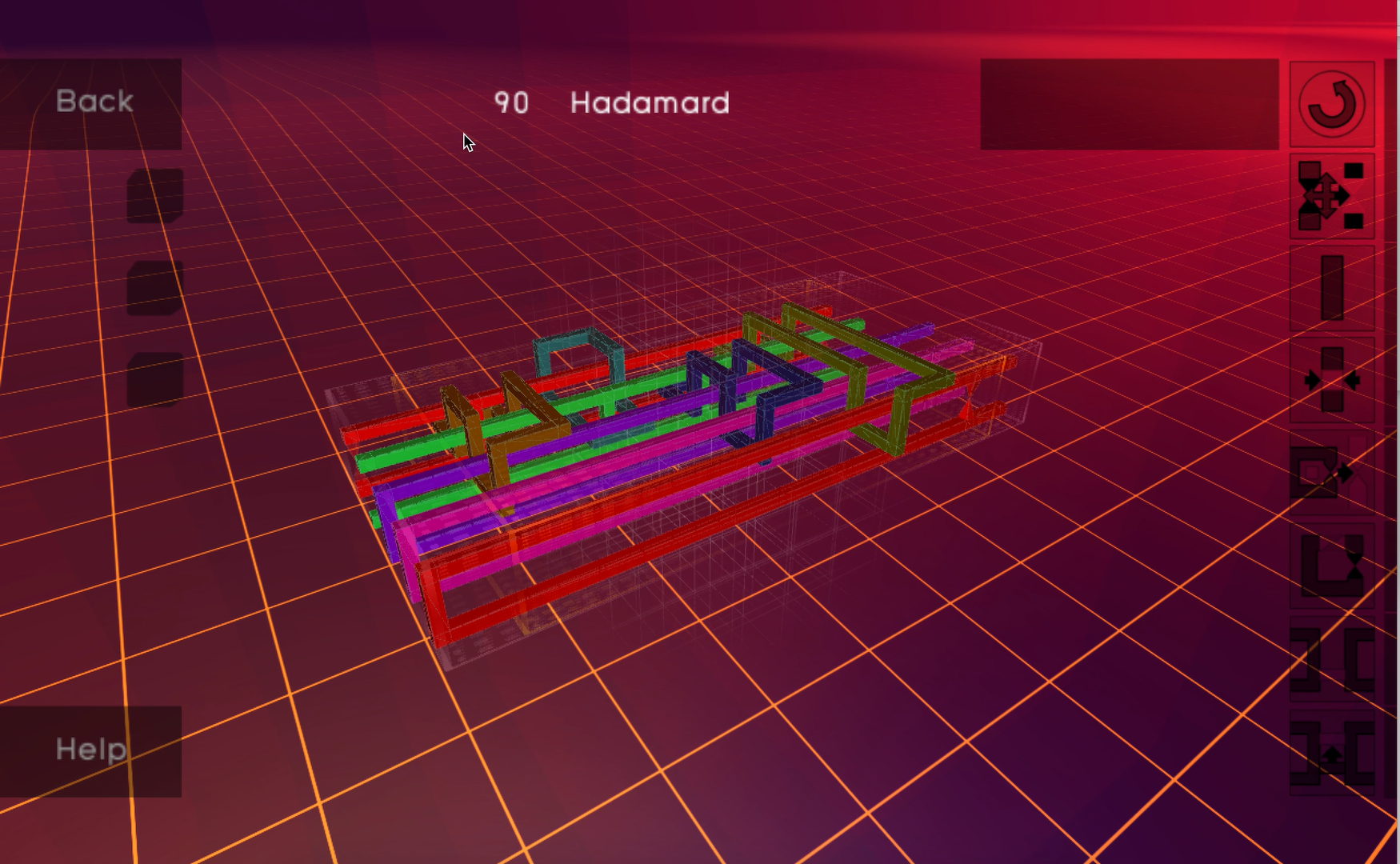}}\\
\caption{(a) An annotated interface of the \textit{Game mode} of \textit{Quantum Moves 2} (2019) (Shaeema Zaman Ahmed \cite{ahmed2021PHD}), and (b) screenshots from \textit{meQuanics} (2016). On \textit{Quantum Moves 2}, the player controls the position of a wave-like potential confining a liquid-like quantum object and aims to move this liquid to a designated position. In \textit{meQuanics}, the player solves puzzles consisting of complex knot-like structures. \textit{Quantum Moves 2} and \textit{meQuanics} are citizen science games for quantum physical sciences.}
\label{screenshotsqm}
\end{figure*}

Another motivation to incorporate numerical simulations of quantum physical phenomena into games lies in the use of citizen science \cite{piispanen2024cisci, piispanen2024thesis}. Quantum physics-related games have been developed for data gathering and problem solving in quantum sciences \cite{piispanen2024cisci}; \textit{Quantum Moves} (2012), \textit{Alice Challenge} (2016) and their successor \textit{Quantum Moves 2} (2019) \cite{alice, jensen2021} are citizen science games developed by \textit{Science at Home}, where players find solutions meant to optimise a certain quantum state-transfer process within the framework of quantum optimal control \cite{cong2014, sklarz2002loading, dowling2003}. In \textit{Quantum Moves 2}, for example, the player attempts to move a 2-dimensional graph with a well-shaped confinement, inside which the visualisation of a numerical quantum simulation sloshes accordingly in a water-like manner (see Figure \ref{screenshotsqm}a). This corresponds to a procedure done with lasers in the laboratory. %The aim of the game is to position this “quantum liquid” at a designated spot. 
Other research groups have attempted to test the fundamental theory of quantum information itself through a game called \textit{The Big Bell Test} (2016) \cite{bellgame2018} and developed \textit{meQuanics} in 2016 as a prototype citizen science game for optimising quantum algorithms \cite{devitt2013,devitt2016}. In \textit{meQuanics} the player aims to unravel complex knot-like structures according to the rules given in the game (see Figure \ref{screenshotsqm}b). Games have also been developed to allow the public to design methods for quantum error correction simply by solving puzzles, as in the game \textit{Decodoku} \cite{wootton2017}.

Numerical quantum simulations have also been turned into interactive visual tools by projects such as \textit{Spin Drops} \cite{spindrops}, the physical art piece \textit{Quantum Garden} and its descendants, the virtual reality simulation \textit{Quantum Playground}, and the large-scale version of \textit{Quantum Garden}, called \textit{Quantum Jungle} \cite{qgarden, qplaygroundGL, piispanen2024thesis, qjungle}. Although not presented as games, the latter three incorporate gamified elements, such as set goals and a rewards, and can therefore also be considered as games \cite{piispanen2024thesis}. The user initiates a numerical simulation of the probability distribution of a particle evolving through a network as a colourful wave through the art piece, then tries to guess where the particle ended up by touching the art piece. A colourful animation rewards the observer's success. 

\subsection{Games on Quantum Computers}
\label{sec:onqcomp}
Through games, it was possible to explore the limits of early computers and to test experimental setups inspired by game theory. Similar motivations have driven the development of the first \textit{quantum computer games} — games designed to run on quantum computers \cite{wootton2018, cantwell2019quantum, wu2021board,perez2024game}. To support this, actors like IBM Quantum, Google AI as well as Microsoft, have developed low-level and accessible tutorials with playful elements, including instructions on developing games and other programs on quantum computers \cite{woottonHelloQuantum, katas, quantumai}.

The first trials of games on quantum computers were made in 2017 on the first publicly accessible device by IBM Quantum, using the newly released system development kit, \textit{Project Q}. Adapted to suit the strength of the qubits available at the time, the game \textit{Cat/Box/Scissors}, inspired by \textit{Rock-Paper-Scissors}, became the very first game to run on a quantum computer \cite{wootton2017cat,wootton2018history,wootton2022,piispanen2023history}. This game is essentially a quantum version of \textit{Rock-Paper-Scissors}, with quantum-sourced randomness determining the opponent's moves. It allows for the possibility of developing a winning strategy against the noise-induced randomness in the opponent's moves. 

Shortly after, a multiplayer game called \textit{Quantum Battleship} was developed on a quantum computer \cite{wootton2022}. It was designed to embody decoherence affecting the quantum phenomena of entanglement between two qubits, as the qubits used had a considerable amount of noise in them. Both games were command-line-based and required lengthy wait times for processing moves due to the long queues on quantum computers.

\textit{Quantum Solitaire} (2017) is an online card game connected to a quantum computer, developed using Unity \cite{wootton2018history,woottonsolitaire}. A quantum computer generates certain types of data needed for the game beforehand, making \textit{Quantum Solitaire} faster and more responsive than earlier quantum computer games. Although randomness in \textit{Quantum Solitaire} is generated conventionally, it is based on probabilities calculated by a prototype quantum computer. Thus, while the game does not access a quantum computer directly during gameplay, its mechanics are governed by quantum phenomena. In this game, the player aims to collect all the red and black cards from the deck using as few moves as possible. Knowledge of quantum mechanics can be used to determine the best strategies for success.\\

Educational games about quantum computers have accompanied the release of open-access cloud services. In 2018, a two-player cooperative board game, \textit{Entanglion}, was created at IBM Research to introduce basic quantum computing concepts \cite{weisz2018}. The game, available open source, introduces players to various aspects of quantum computing, including hardware and software components, through a narrative set in a futuristic galactic adventure \cite{entanglion}. The mobile game \textit{Hello Quantum} (2018) was developed in collaboration between IBM Quantum and the University of Basel for teaching the logic behind quantum computational operations through approachable puzzles \cite{seskir2022}. Educational materials titled \textit{Hello Qiskit}, were also included as part of IBM Quantum’s interactive textbook \cite{wootton2020teach}. \textit{$Q| Cards >$} (2019) was created as an online card game for teaching basic logic behind quantum computing and determining the winner on a quantum computer \cite{kultima2021qgj, wootton2022}. It was developed originally at the game developing event \textit{Quantum Game Jam} (introduced in more detail in the next subsection) and further developed together with IBM.\\

Quantum computers have not only been used as hardware for games, but their development and the advantages of quantum algorithms have also inspired level design in games. Procedural map generation that combines quantum computer-generated randomness with algorithms has shown promising results \cite{wootton2020,wootton2020a}. These proof-of-principle methods are based on the idea of using quantum interference to create unique effects in a blurring process. The concept has been generalised for use in procedural generation for music and art, as well as in computer games \cite{wootton2021}. 

\subsection{The Rise of Quantum Game Development} 
Until 2014, quantum physics-related games presented as “quantum games” were primarily produced by research groups composed mainly of quantum physicists and experts. That year marked the organisation of the first \textit{Quantum Game Jam}, which led to a series of five quantum-themed \textit{game jam} events that brought together individuals from both the quantum physics research community and professional game development to develop quantum physics related games \cite{kultima2021qgj}. In 2019, the event organisers partnered with IBM Research and IBM Quantum in order to provide more mentoring and dedicated access to quantum computers. Since then, other actors like the Indian community for Quantum Computing, IndiQ, and the Quantum AI Foundation have also organised quantum computing-themed game jams and hackathons \cite{indiq,quantumai}. 

Like most game jams and hackathons, \textit{Quantum Game Jams} were forced to go online during the COVID-19 pandemic, as seen in October 2021, when an online \textit{Quantum Game Jam} was held in collaboration with Aalto University, the University of Turku, and IGDA Finland \cite{piispanen2023qgj}. To date, at least 160 games have emerged from quantum physics-themed game jams, along with over 60 quantum physics-related games from hackathons and university courses like the \textit{Aalto Quantum Games} course \cite{quantumgames, piispanen2022course}. Many of these games have been developed using quantum computing software such as Qiskit, offered by IBM Quantum, and some even utilise actual quantum computers. These game prototypes are often designed either to educate on quantum physics-related concepts or to serve as citizen science games.\\

The series of \textit{Quantum Game Jams} has not only led to the creation of numerous prototypes for science games in the field of quantum physics \cite{kultima2021qgj,piispanen2023projects, piispanen2024thesis} but also fostered the development of creative projects related to quantum physics, including commissioned artwork like \textit{Quantum Garden} and the VR-installation \textit{Quantum Playground} \cite{qgarden,qplaygroundGL,piispanen2023projects, piispanen2024thesis}. Quantum physics themed game jams have thus significantly influenced the definition of quantum games as presented in this article. Additionally, the development of quantum physics-related games has provided opportunities for the creators of these games to learn about quantum physics, quantum technologies, and quantum computing.

\section{Characterising the Dimensions of Quantum Games} 
\label{sec:dimensions}
Gordon and Gordon defined \textit{quantum computer games} as \textit{“computer games where the rules of the game are based on quantum principles, such as superposition, entanglement and the collapse of the wave function”} \cite{gordon2010}. They introduced this description to distinguish their ‘quantum computer game’, \textit{Quantum Minesweeper}, and computer games alike from \textit{quantum game theory}, a formal mathematical theory that incorporates quantum strategies into classical game theory\footnote{Also known as interactive decision theory, the theory surrounding optimal strategies related to decision making.} \cite{gordon2010}. Although this definition was initially intended for a specific educational tool, it has since been cited as the definition of \textit{quantum games}. More recently the term \textit{quantum game} has been referred to as “\textit{computer (or video) games with one or more of the phenomena from quantum physics embedded in their game mechanics}” to refer to a set of quantum physics-related computer games designed for educational purposes \cite{seskir2022}. 

Given the recent increase in games that refer to or mirror quantum physical phenomena in a more inspirational way, as well as the rise of analogue games referencing quantum physics and games running on quantum computers, these definitions are no longer sufficient. Quantum computers not only allow the design of circuits to test game-theoretic quantum strategies \cite{du2002}, but they also offer much more for games. The connection between a game and quantum physics can be complex, and this article aims to bring clarity to what quantum physics-related games are.\\

Before undertaking this study, the authors each had their own approach for distinguishing quantum physics-related games based on their purposes and the ways they employed quantum technologies or numerical simulations. These approaches were not formal classifications and had not been tested or validated against existing quantum physics-related games. These classifications were also not published, but used in presentations related to outreach events and educational lectures. While these classifications shared similarities, they were not identical. In particular, the games from \textit{Quantum Game Jam}s challenged the previous definitions of quantum games and highlighted the need for new categorisations. Some games did not have any perceivable elements or materials directly related to quantum physics, even though they were developed under a quantum physics-related theme with active involvement from quantum physics researchers. Other games, despite being primarily inspirational, still used numerical simulations of quantum physical systems for various purposes. To facilitate discussions about these games from the perspective of game development, the authors examined the quantum physics-related games known to them to identify their most distinctive attributes. Based on the examination of over 250 quantum physics-related games known to the authors at the time of this analysis carried out between 2019 and 2022, we propose a three-dimensional framework to better define and analyse quantum games. 

\subsection{Methods}
\label{sec:methods}
We base our definition on an analysis of existing quantum physics-related games and the authors' years of experience in developing them. The authors came together in late 2019 to discuss various aspects of quantum physics-related games and their connections to quantum technologies, quantum education, and numerical simulations depicting quantum mechanical behaviour. We collected and analysed an extensive set of quantum physics-related games to identify the distinctive aspects of their relationship to quantum physics.\\ 

The collection of quantum physics-related games has been gathered between 2018 and 2022 by searching through online databases like itch.io, mobygames.com, Steam and igdb.com using search terms like “quantum”, “quantum physics”, “particle”, “superposition”, “entanglement”, “multiverse”, and other quantum physics-related keywords. In addition, other online repositories, conference proceedings, and academic journals were searched using search engines like Google Scholar, Google, and Ecosia, as well as integrated search engines of databases such as the ACM Digital Library, IEEE Conference Proceedings, and online preprinting services like arXiv.org and TechRxiv.org. These searches focused on publications related to quantum physics, serious games about quantum physics, citizen science games about quantum physics, and events like the \textit{Quantum Game Jam} \cite{kultima2021qgj, piispanen2023qgj}, \textit{IndiQ Quantum Game Jam} and various hackathons. These searches continued until August 2024 repeatedly as a way to keep track of most recent additions (like the important addition of \textit{Quandoom}, an adaptation of the game \textit{DOOM} by Id Software in 1997 for a quantum computer \cite{quandoom}), but the analysis presented here included only games that were publicly available between the years 2018 and 2022.

The initial collection of games was conducted by the first author from academic sources and online databases between 2018 and 2019 and compiled into a local LaTeX document that was openly circulated as a PDF between 2019 and 2021 to anyone interested in quantum physics-related games. This led to further contacts and some additions to the list. Several important additions were incorporated into this list from a QIntern project in 2021 organised by the global quantum technology network QWorld Association \cite{qintern}; their project used the aforementioned PDF as one of their main sources \cite{garcia2021}. The first author compiled all the aforementioned games into a single list in an online spreadsheet, tagged according to educational or citizen science purposes, the use of numerical simulations of quantum physical systems, the use of quantum software or hardware, or credits to quantum physicists as developers. In October 2021, this list was shared with authors MP, JW and AK for discussion. 

A public copy of the list, containing only the main information about these games and not the tags, has been openly available with contact information since March 2022. This version of the list has been actively shared and been open for suggestions \cite{quantumgames}. Consequently, the authors have been able to include information and online repositories of games from courses or events not otherwise published and test their proposal on newer additions. Games have been added to this list by the first author until the publication of this article, and the list remains open for future additions \cite{quantumgames}. In addition, a preprint of this analysis has been available on the preprint services arXiv.org since 2022 and TechRxiv.org since 2023, which allowed the authors to track the existence of later publications on quantum physics-related games.\\

Concurrent meetings were held alongside the collection of games starting in October 2021, which provided a platform to discuss newly emergent aspects of the games. These discussions sought to find a structure for analysing and describing the listed games, and included topics such as determining whether it was relevant to separate the use of numerical simulations describing quantum physical phenomena (usually developed for citizen science use \cite{piispanen2024cisci,piispanen2024thesis}), quantum computing software, and quantum hardware and in what way. Project-dependent numerical simulations, such as the \textit{Quantum Black Box} used for several \textit{Quantum Game Jam} prototypes of citizen science games, were decided to be kept separate from the rest, as they were most often connected to a serious purpose and thus individualised in the analysis \cite{qbb, kultima2021qgj,piispanen2023projects, piispanen2024cisci, piispanen2024thesis}. 

To identify suitable characterisations of quantum physics-related games, the study aimed to answer the following questions:
Has a quantum computer been used in developing the game, or is a quantum computer being used for playing the game? Have numerical simulations based on quantum mechanical calculus or quantum computing software (like Qiskit from IBM, Q\# by Microsoft or Cirq by Google AI) been used? What has been the purpose for introducing quantum physics to the game (Educational, Citizen Science etc.)? How is quantum physics related to what is perceivable in the game?

For this study, over 250 games were studied from the \textit{List of Quantum Games} between the years 2019 and 2022 \cite{quantumgames}. The games went through an iterative analysis ran by the first author and guided by the above questions in a manner that inspires from transcript coding \cite{cope2010}. For the perceivable aspects, separate columns were created for aspects such as “Is a numerical simulation of a quantum physical system visible in the game?”, “Are there references to quantum physics in the narrative?”, or “Are there literal visual representations of quantum physical phenomena or technologies?” Whenever an aspect of the game related to quantum physics could not be described using the existing questions, a new column was added to accommodate this aspect, and the list of games was reviewed for similar characteristics.  

As the number of aspects grew, a grouping for them was reformed to separate perceivable aspects from the use of quantum technologies or numerical simulations and to identify the possible serious use of the game. For example, purposes related to teaching different aspects of quantum computing or quantum theory were grouped under “educational purposes”, with a column added to list aspects like “Concepts you learn about, and are explained correctly” inside a single cell separately from “Concepts you hear/see being said and used, not explained necessarily”. The final set of questions is shared in table \ref{table:finalquestions}. In addition to the questions listed, information was also gathered about the creators, the platform the game can be played on, the software used for developing the game, the country that the game originates from, the year the game has been published, and the possible event the game relates to.

The authors took the liberty to fill in information for games with no documentation when they had contacted the developers, were involved in the development process, or otherwise acquired the information. Initially, the analysis focused on aspects perceivable from the gameplay experience, but it was acknowledged that peripheral materials related to games (such as rule books, developers' descriptions, release notes, press kits, development event details, etc.) are an essential part of the games. In cases where the term “quantum” was used merely as a loose sci-fi reference and no other connections to quantum physics were found, these games were not tagged under any of the three aspects and removed from the list.

By the end of the analysis ran by the first author and the discussions together with all the authors, three distinctive aspects were found to constitute a quantum game, that corresponded to any perceivable aspects of quantum physics in the game, to the use of quantum technologies, like quantum software or quantum hardware, and to the serious purposes of the games. For a final testing the suitableness of these aspects, a simple Yes/No table was conducted by the first author on the list of games to answer if a game coincided with the description of a particular set of aspects. In Table \ref{table:examples}, we present an example listing of a few of the games introduced in Section \ref{qpgames} for testing the dimensions. A more extensive listing of the games discussed in this section can be found in Table \ref{table:exampleslong}. As no examples of quantum physics-related games were found without at least one of these aspects, it was concluded by the authors to propose these aspects as the \textit{dimensions of quantum games}. 
\begin{table}[!ht]
	\caption{List of seven example games out of the examined 250 games, characterised using the dimensions of quantum games. For each game, a Yes/No answer indicates whether the game exhibits the qualities described by the three dimensions of quantum games; the perceivable dimension of quantum physics (“Perceivable”), the dimension of quantum technologies (“Q. Tech”), and the dimension of scientific purposes (“Scientific purpose”) related to quantum physical sciences. In the “Perceivable” dimension, “simulation” is noted for games that numerically simulate perceivable quantum physical phenomena on a classical computer. In the final column, the specific scientific purposes of each game are indicated in parentheses. “Citizen Sci” refers to Citizen Science initiatives and “Benchm” refers to benchmarking early quantum hardware.}
	\label{table:examples}
 \setlength{\tabcolsep}{6pt} % Default value: 6pt
\renewcommand{\arraystretch}{1.3} % Default value: 1
	\centering
\begin{tabular}{|llll|}
\hline
Game                     & Perceivable  & Q. Tech & Scientific purpose  \\
\toprule
\hline
Quantum TiqTaqToe        & Yes (simulation)   & No	& Yes (Education)   \\
Quantum Game  	         & Yes (simulation)   & No 	& Yes (Education)   \\
Cat/Box/Scissors	     & Yes 	 & Yes  & No \\
meQuanics		         & Yes (simulation)	 & No	& Yes (Citizen Sci)  \\
Quantum Awesomeness	     & Yes	 & Yes	& Yes (Benchm)  \\
Quantum Break	         & Yes   & No   & No  \\
C.L.A.Y.		         & No    & Yes  & No \\ \hline  
\end{tabular}
\end{table}

\subsection{The Three Dimensions of Quantum Games}
Quantum physics can be incorporated into a game using several different methods. Therefore, strict categories may limit the exploration of potential connections a game might have to quantum physics. Instead of providing strict categories, the concept of \textit{dimensions of quantum games} is suggested and used by the authors. What became particularly clear and motivated the authors to adopt a broader definition of quantum games was that there were now games that were deeply connected to quantum physics by running on a quantum computer, though the rules of the game might not have anything to do with quantum concepts. These games generally do not fit the early definitions for either quantum games or science games. 

In our study, we have found that games may reference or relate to quantum physics, quantum technologies or quantum computing through three distinct aspects, which we refer to as the \textit{dimensions of quantum games}: the perceivable dimension of quantum physics, the dimension of quantum technologies, and the dimension of scientific purposes (related to quantum physical sciences).

\subsubsection{The Perceivable Dimension of Quantum Physics}
In the \textbf{perceivable dimension of quantum physics}, “perceivable” refers to awareness of quantum physics or quantum physics-related concepts through the various layers of the game piece. This includes, for instance, the graphical, narrative, and thematic elements of the game, along with the mechanics and the rules of the game. \textit{The reference to quantum physics in the game is perceivable by interacting with the game or with its peripheral material (such as rule books, developers' descriptions, etc.)}. 

\begin{figure*}[ht]
\center
\subfloat[]{\includegraphics[width=0.30\linewidth]{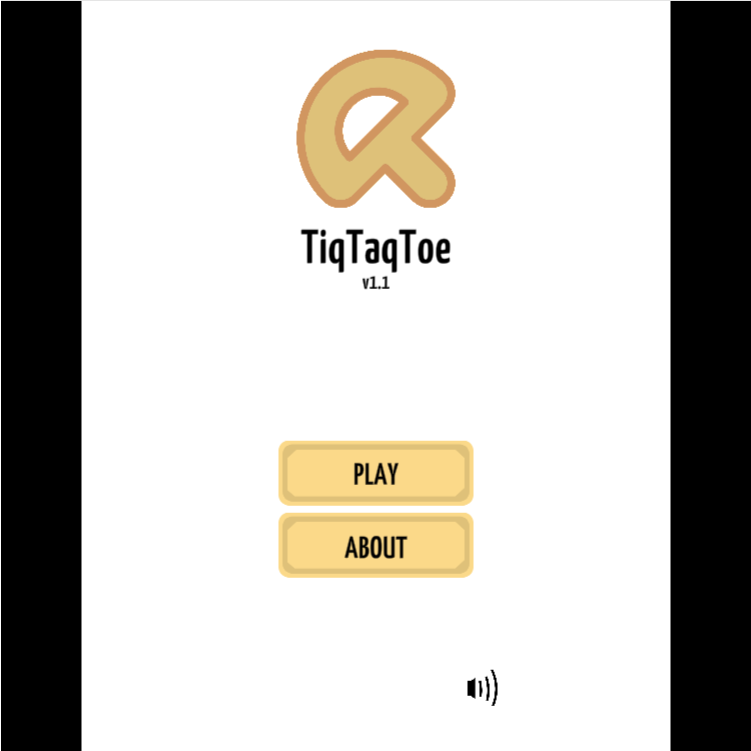}
\includegraphics[width=0.30\linewidth]{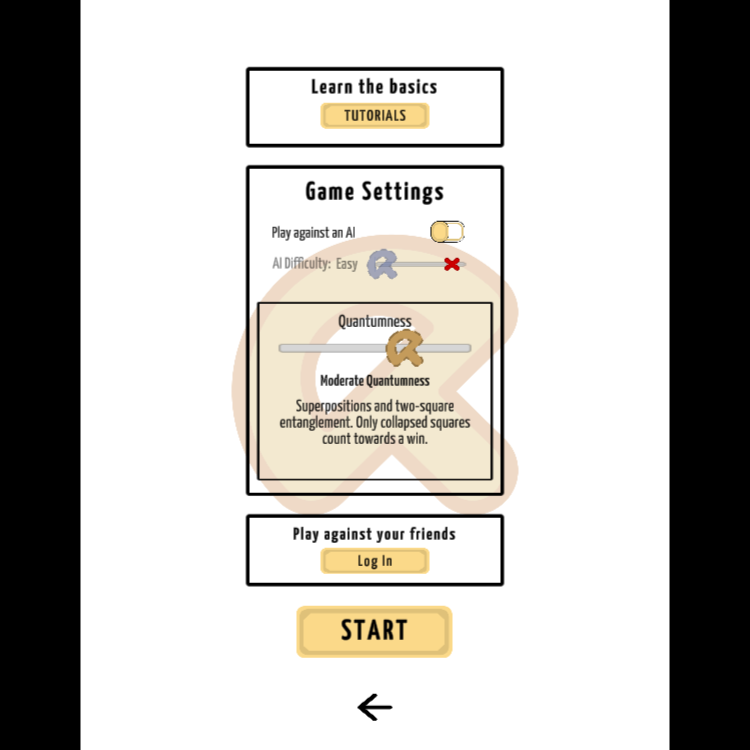}
\includegraphics[width=0.30\linewidth]{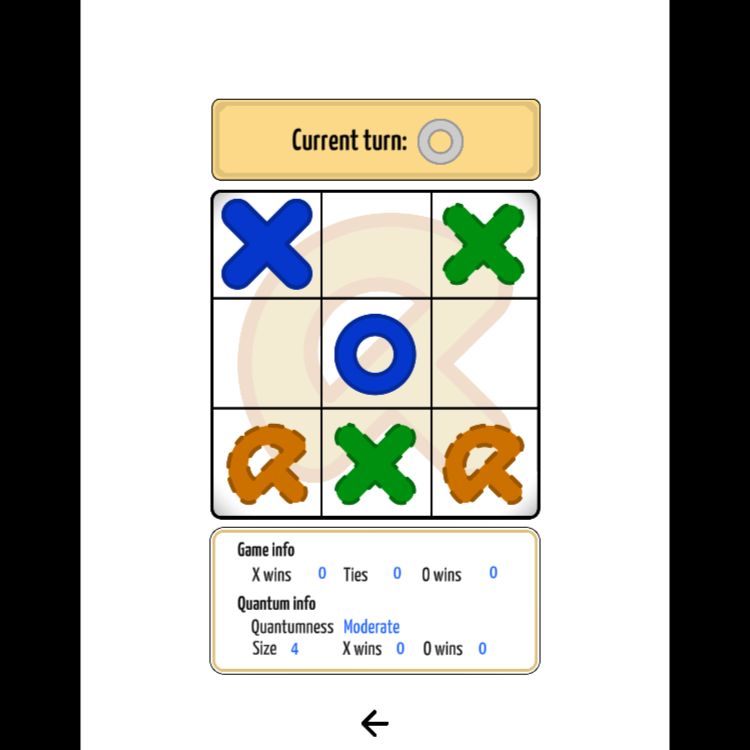}}\\
\subfloat[]{\includegraphics[width=0.45\linewidth]{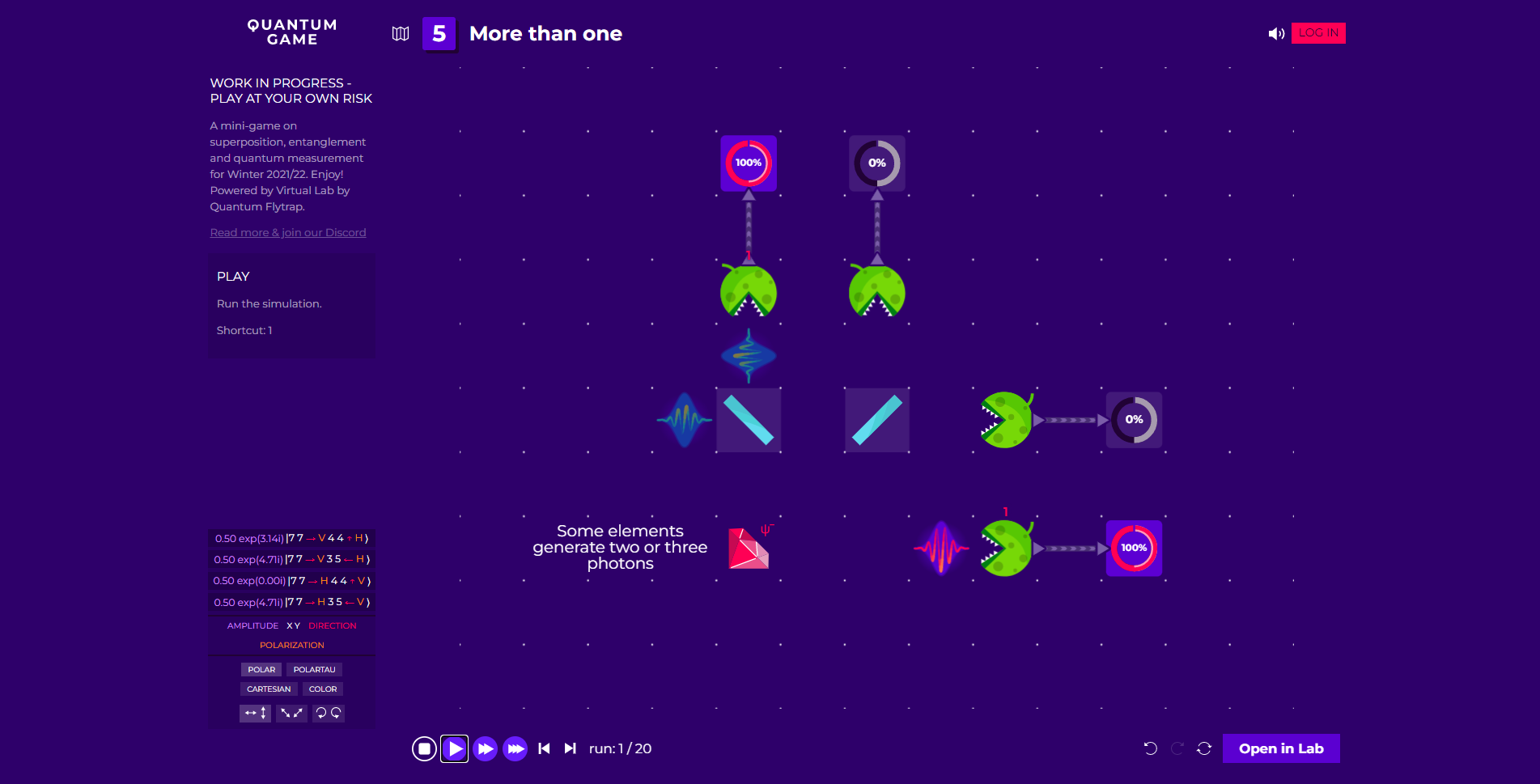}\,
\includegraphics[width=0.45\linewidth]{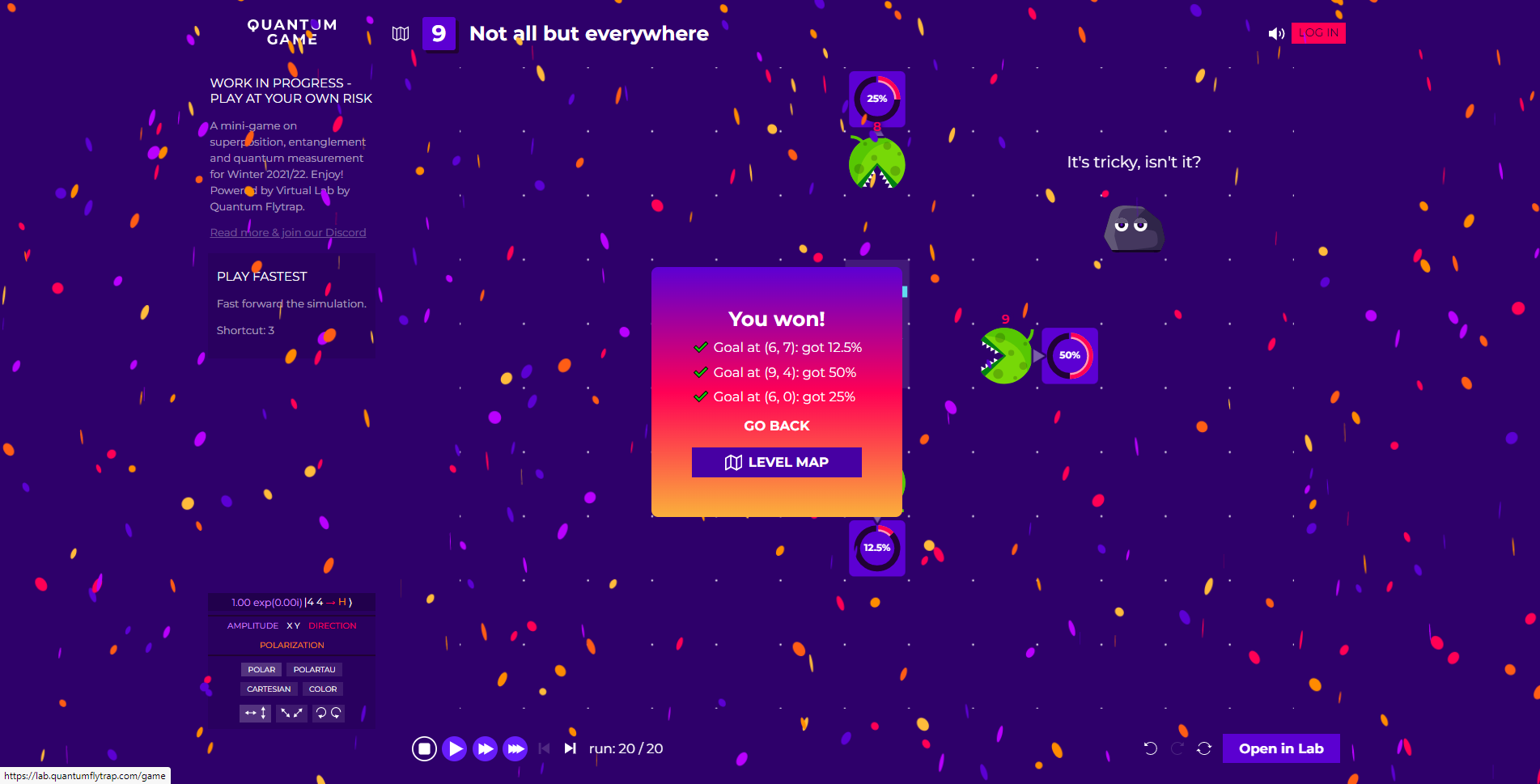}}\,
\caption{Screenshots from the educational games (a) \textit{Quantum TiqTaqToe} (2019) and (b) \textit{Quantum Game} by Quantum Flytrap (2020). In \textit{Quantum TiqTaqToe} the basic rules of the classic Tic-Tac-Toe are enhanced with actions displaying quantum mechanical phenomena. In \textit{Quantum Game} the player solves puzzles introducing the basic actions and logic of quantum optics equipment.}
\label{screenshots00}
\end{figure*}
For a game to have a strong perceivable dimension of quantum physics would mean that the game depicts a reference to quantum physics in multiple, noticeable ways. The game can, for instance, feature a visual representation of a numerical simulation designed to replicate quantum mechanical behaviour, as seen in \textit{Quantum Moves 2} introduced in Section \ref{simul} or \textit{Hamsterwave}, where the probability distribution of a one-dimensional Bose-Einstein condensate is portrayed as an evolving wave that the player must reshape to save a hamster sailing on it \cite{piispanen2023projects, piispanen2024thesis}. A game can also have clear rules based on quantum physical principles, like in the game \textit{Quantum TiqTaqToe} (2019), which combines quantum mechanical dependencies with the classical game of \textit{Tic-Tac-Toe} and teaches the player about the logic behind them in its tutorial (see Figure \ref{screenshots00}a) \cite{nieuwenburg2019}. Quantum variations of traditional, classical games like Chess, Checkers, and Poker, for example, have been developed in a way where quantum mechanical behaviour affects the rules, bringing extra dynamics to the gameplay and even teaching basic concepts of quantum physics \cite{cantwell2019quantum,quantumgames}. 
\textit{Quantum Chess} by Christopher Cantwell was carefully crafted to accompany an elaborate mathematical framework for the game and to be inherently quantum, not to educate about quantum physics \textit{per se}, but to allow players to gain intuition about the characteristics of quantum physics \cite{cantwell2019quantum}.

Another type of a perceivable reference to quantum physics is in a game about a quantum physics laboratory like in the game \textit{Quantum Game} by \textit{Quantum Flytrap} (2020), where the player can test the behaviour of laser beams using tools that resemble actual quantum optics lab equipment, and these actions are depicted in a scientifically accurate manner (see Figure \ref{screenshots00}b) \cite{migdal2022visualizing}. All the aforementioned games exhibit a perceivable dimension of quantum physics. 
 
\begin{figure*}[ht]
\center
\subfloat[]{\includegraphics[width=0.30\linewidth]{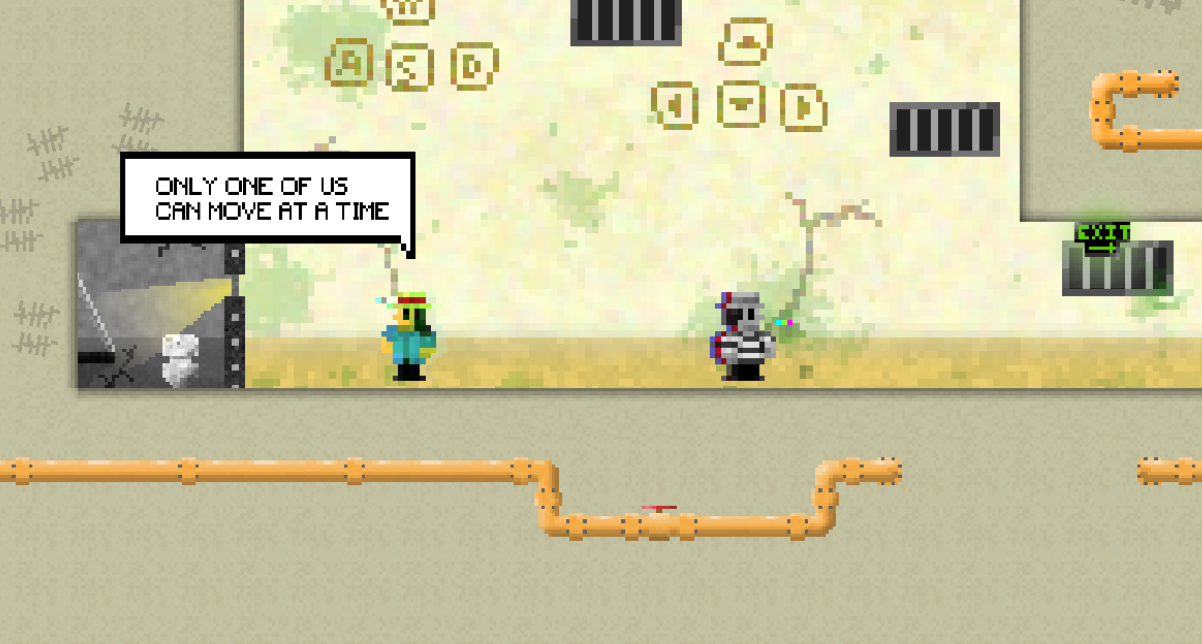}\,
\includegraphics[width=0.30\linewidth]{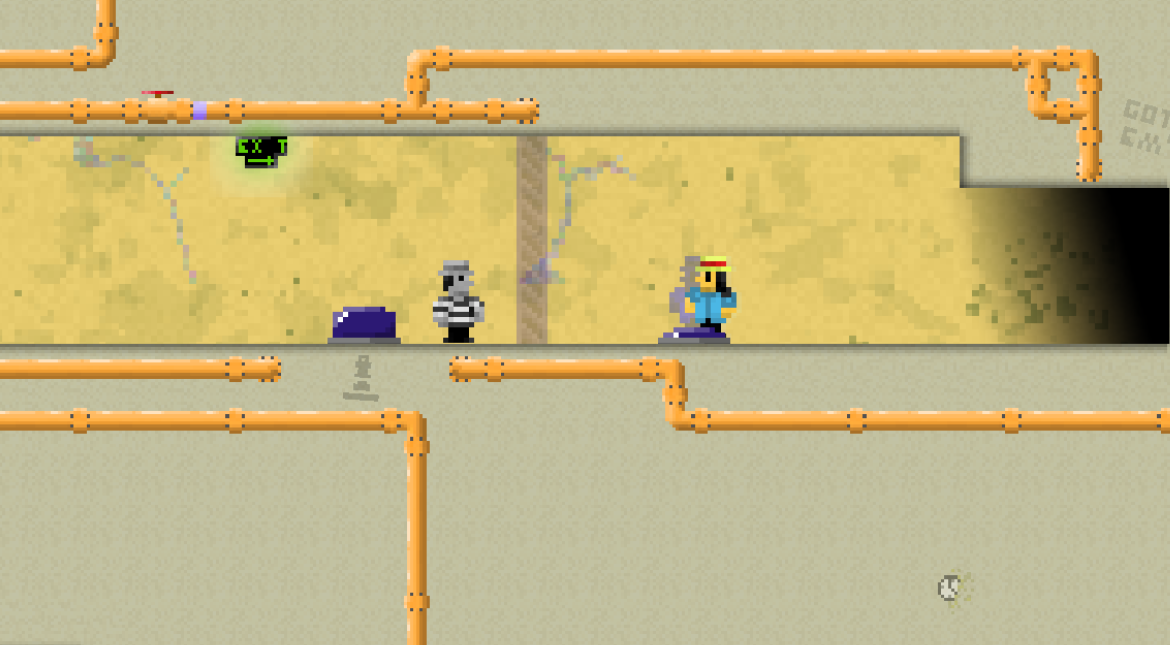}\,
\includegraphics[width=0.30\linewidth]{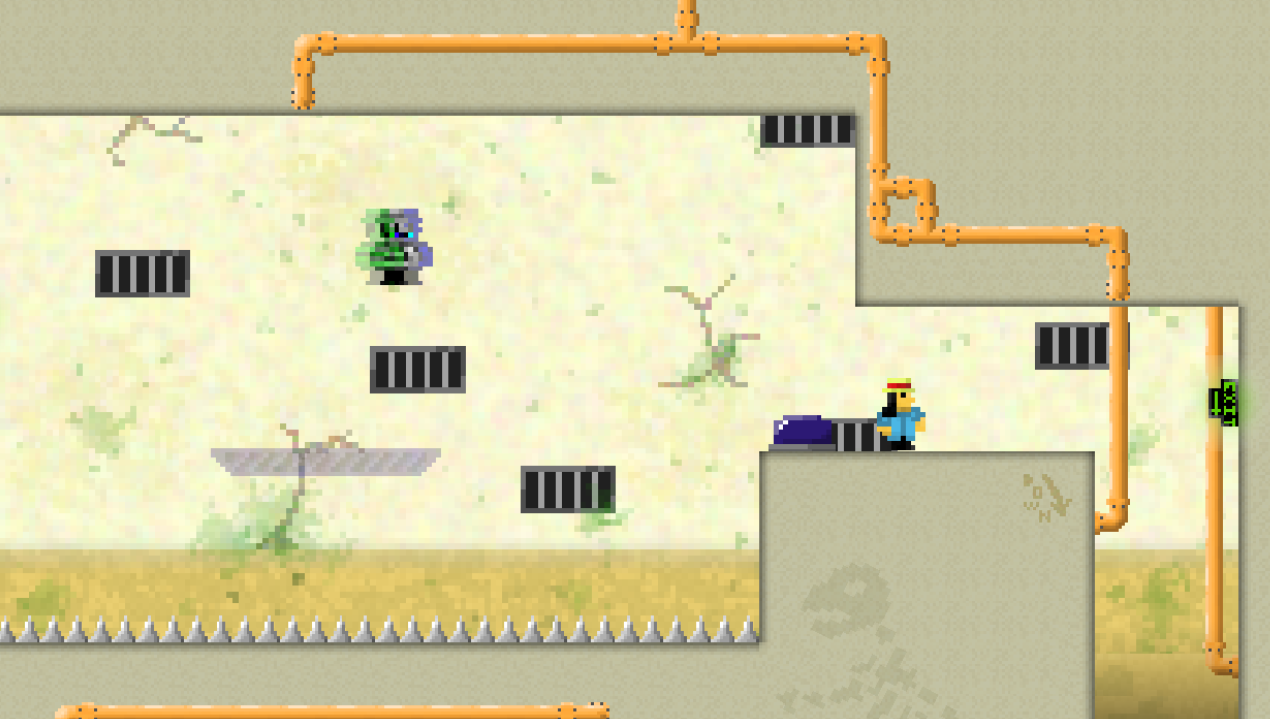}}\,
\subfloat[]{\includegraphics[width=0.30\linewidth]{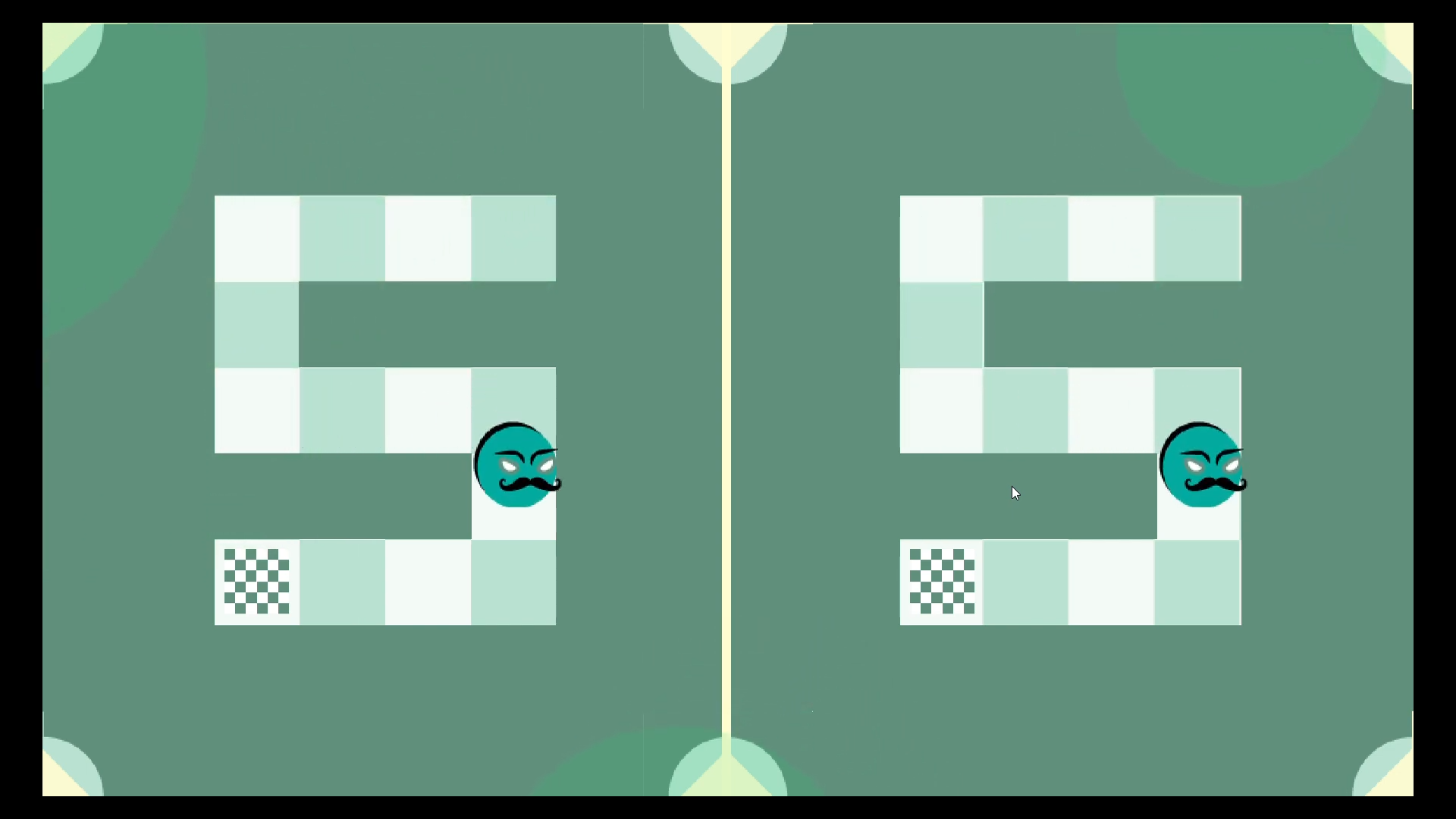}\,
\includegraphics[width=0.30\linewidth]{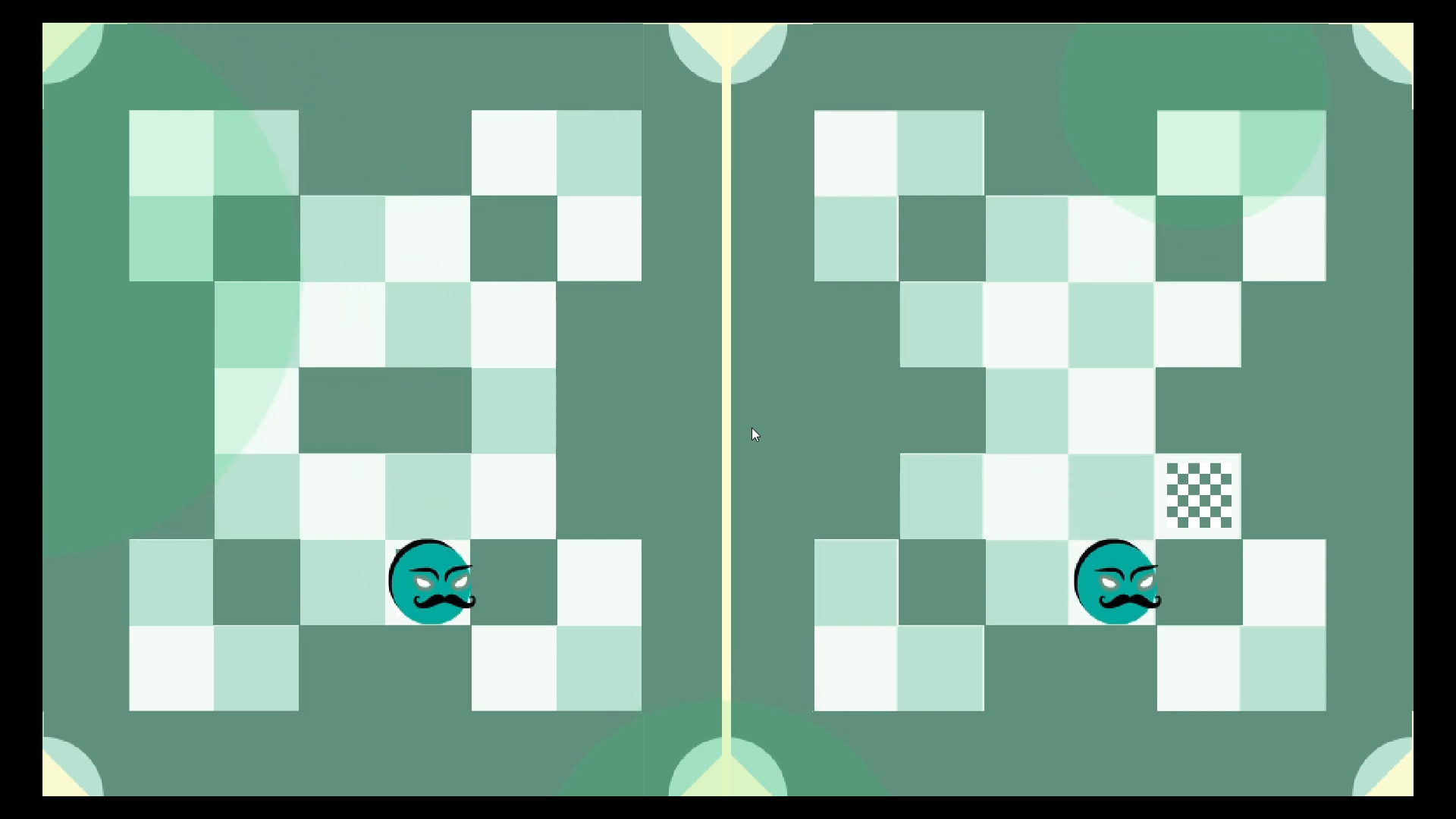}\,
\includegraphics[width=0.30\linewidth]{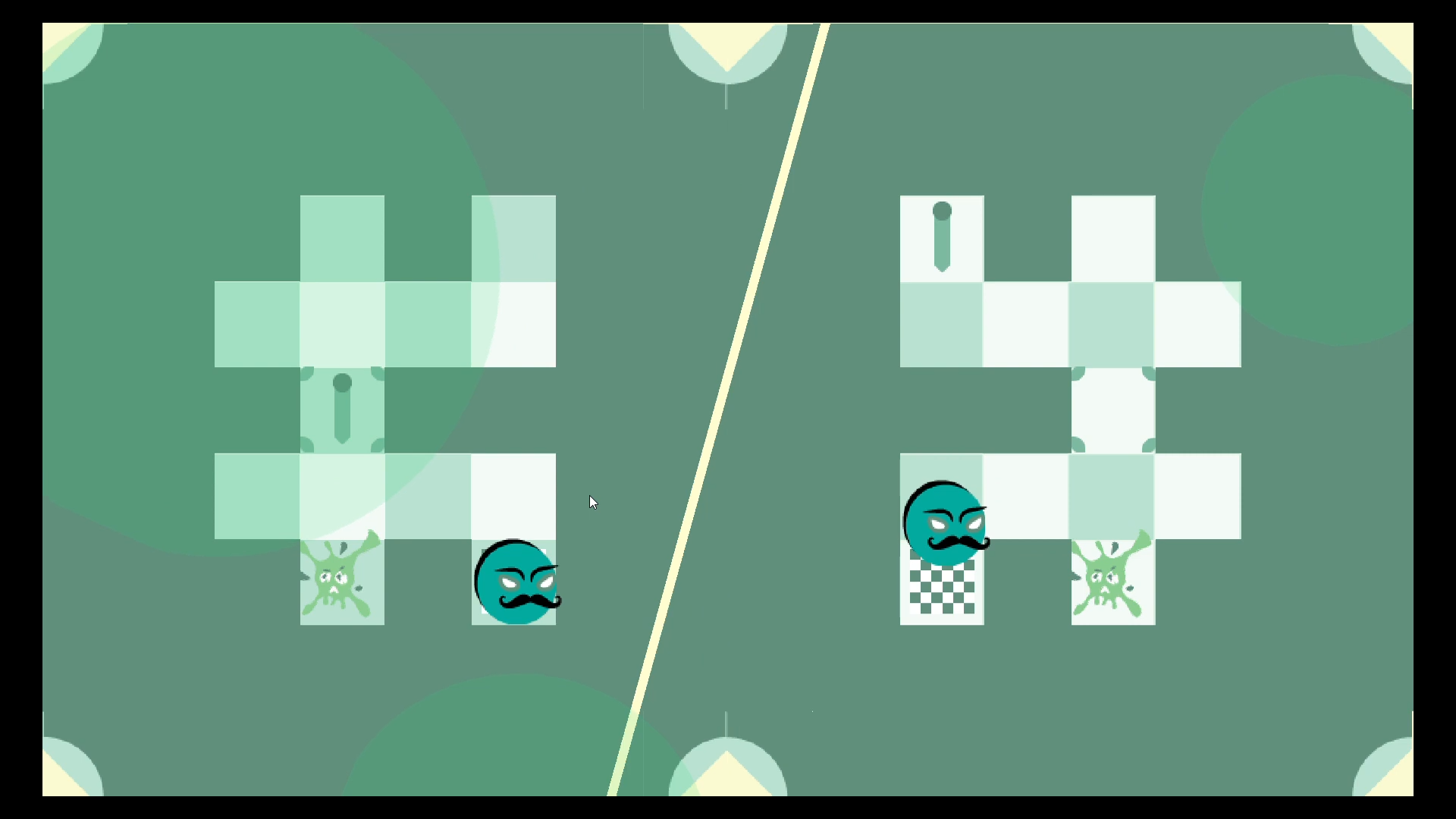}}\,
\caption{Screenshots from games (a) \textit{Quantum Entanglement} (2019), and (b) \textit{Quantum Labyrinth} (2020). \textit{Quantum Entanglement} is a two-player puzzle-adventure, where the player controls two characters, and can only move one of them at a time. In \textit{Quantum Labyrinth} the player characters are controlled simultaneously in order to solve puzzles.}
\label{screenshots01}
\end{figure*}

Examples where elements of the gameplay have been \textit{inspired by} quantum physics but have a slightly weaker connection to it include games where a character appears inherently linked to another, referencing quantum mechanical phenomena like entanglement or superposition. Games like \textit{Quantum Entanglement} (2019) and \textit{Quantum Labyrinth} (2020) depict this characteristic (see Figure \ref{screenshots01}). The shooter game \textit{Escape from Quantum Computer} (2021) references quantum physical phenomena and particles in multiple ways \cite{jaakola2021}. The narrative situates the player inside a quantum computer, where the level design is inspired by the qubit alignment in IBM's \textit{Almanden} quantum computer. Additionally, some character actions refer to tunnelling but have no connection to actual quantum tunnelling through numerical simulation or scientifically accurate means. Instead, they strive to depict the phenomenon through an allegory. A game with a low or disregarded perceivable dimension of quantum physics might use quantum concepts like tunnelling or superposition merely as thematic elements, naming characters or actions without a more meaningful connection to quantum theory. 

A strong example of the perceivable dimension would be a game coinciding with the definition for quantum games given by Gordon and Gordon, but slightly broadened: In a game with a strongly perceivable dimension of quantum physics, \textit{the story-line, developers' descriptions, rules, visuals and actions clearly and correctly reference quantum theory and are based on actual quantum mechanical calculations}.\\

\subsubsection{The Dimension of Quantum Technologies}
We find it relevant to point out that the connection to quantum physics in a game does not always have to be perceivable to be considered a quantum game. A game might not have a perceivable dimension of quantum physics, and yet simulate quantum physical phenomena through quantum software developed with Qiskit, Cirq, Q\# or other quantum frameworks. A game without a perceivable dimension of quantum physics might even run on a quantum computer. Therefore, as the second dimension, we name the \textbf{dimension of quantum technologies} to capture the \textit{use of quantum software or actual quantum hardware either during gameplay or during the development of the game}. 

The first example of a game with a dimension of quantum technologies is the first-ever game to run on an actual quantum computer, \textit{Cat/Box/Scissors} introduced in Section \ref{sec:onqcomp}. The game has a thematic reference to quantum physics through a commonly used \textit{Schrödinger’s cat} reference and the game accompanied an outreach themed educational blog post \cite{wootton2017cat}. This game was designed not only to be played on a quantum computer but also to invoke curiosity about the technology and methodology behind it. We want to point out that this is a feature that will likely be shared by most games implemented on quantum technologies in the coming years, before quantum computers can offer games a definite and convenient advantage. It is for this reason that it can be expected to be known to the player that a game uses quantum devices or quantum simulations, and have it be a possible motivation for playing. 

\begin{figure*}[ht]
\center
\includegraphics[width=0.30\linewidth]{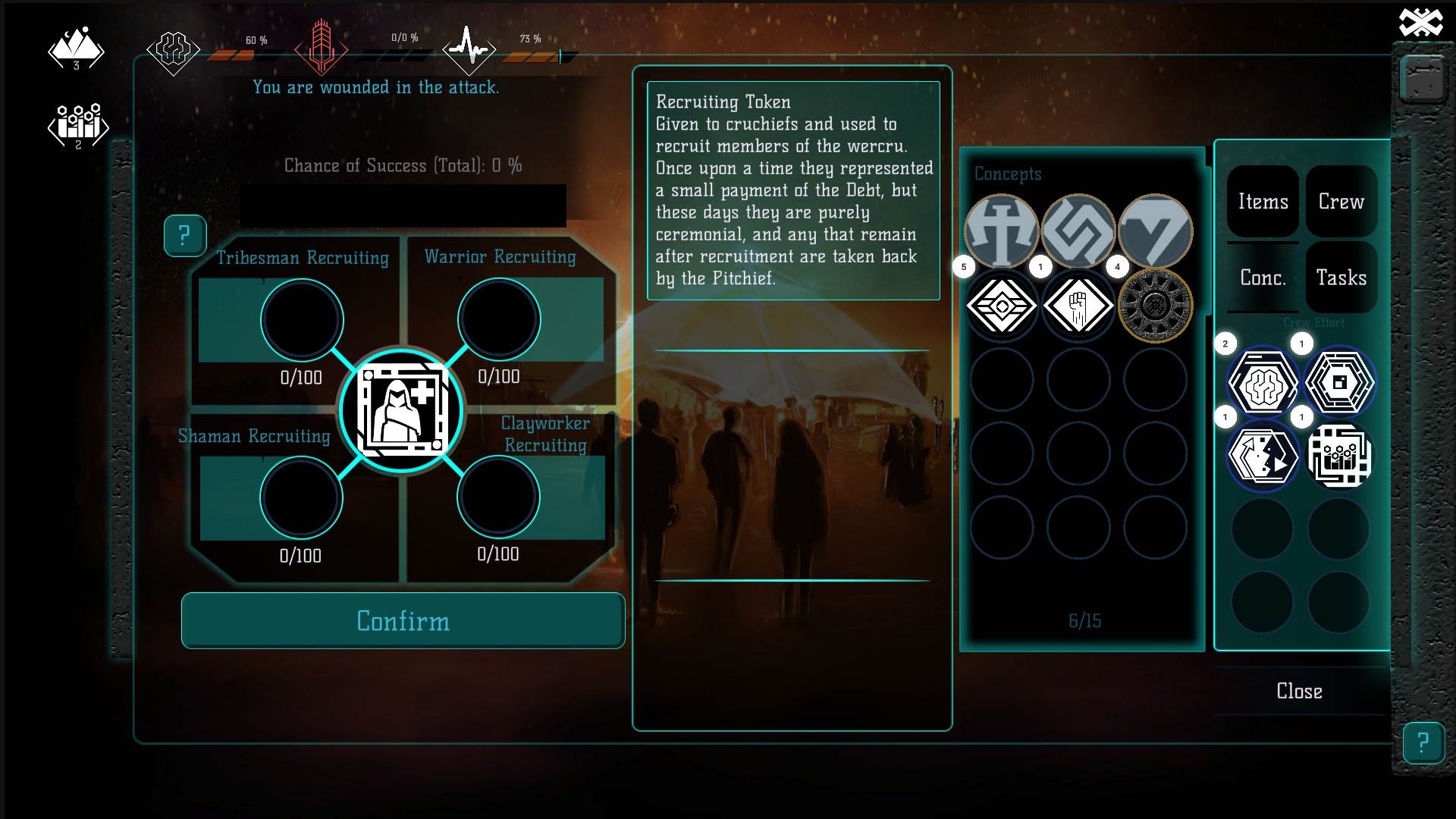}\,
\includegraphics[width=0.30\linewidth]{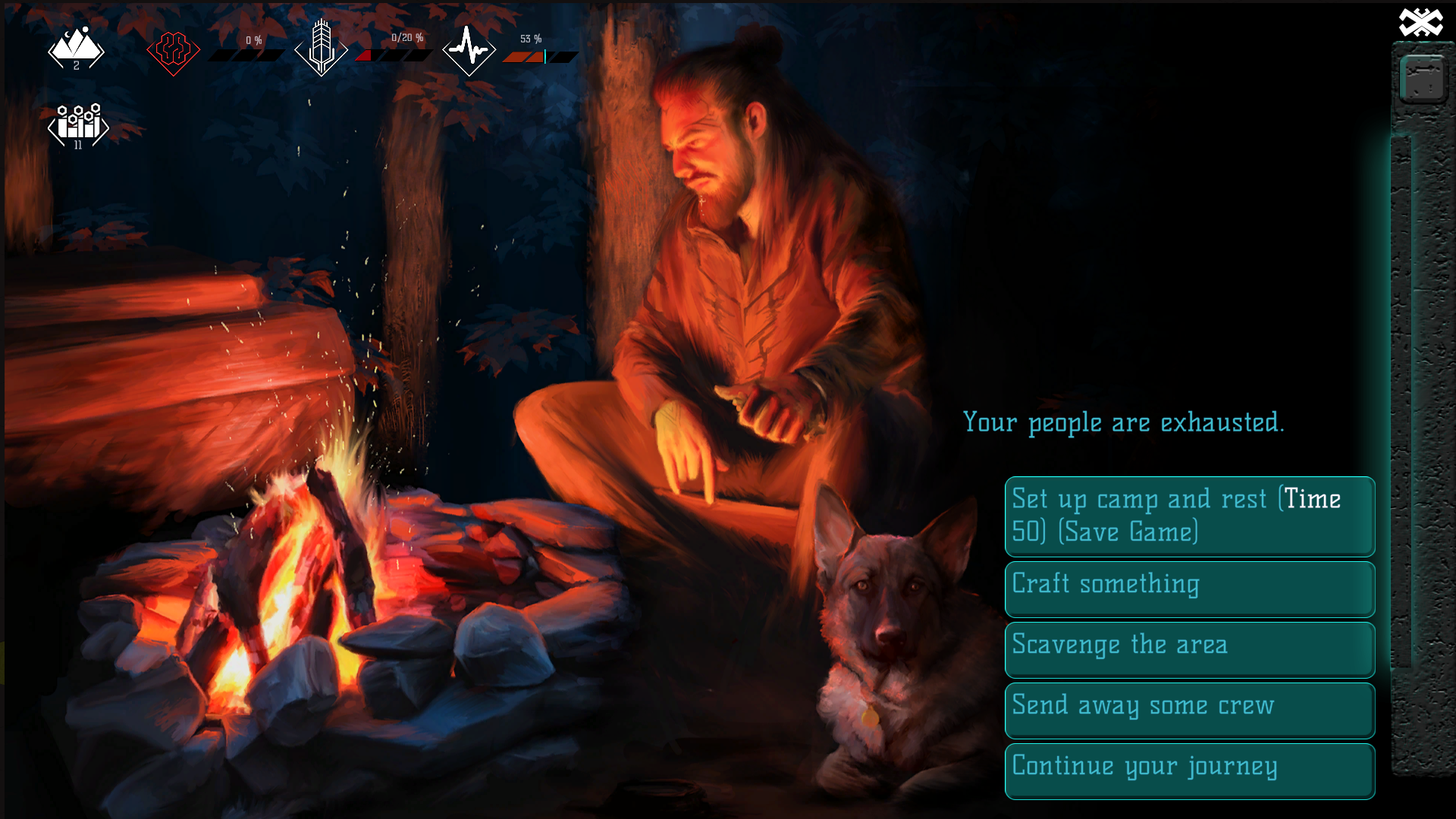}\,
\includegraphics[width=0.30\linewidth]{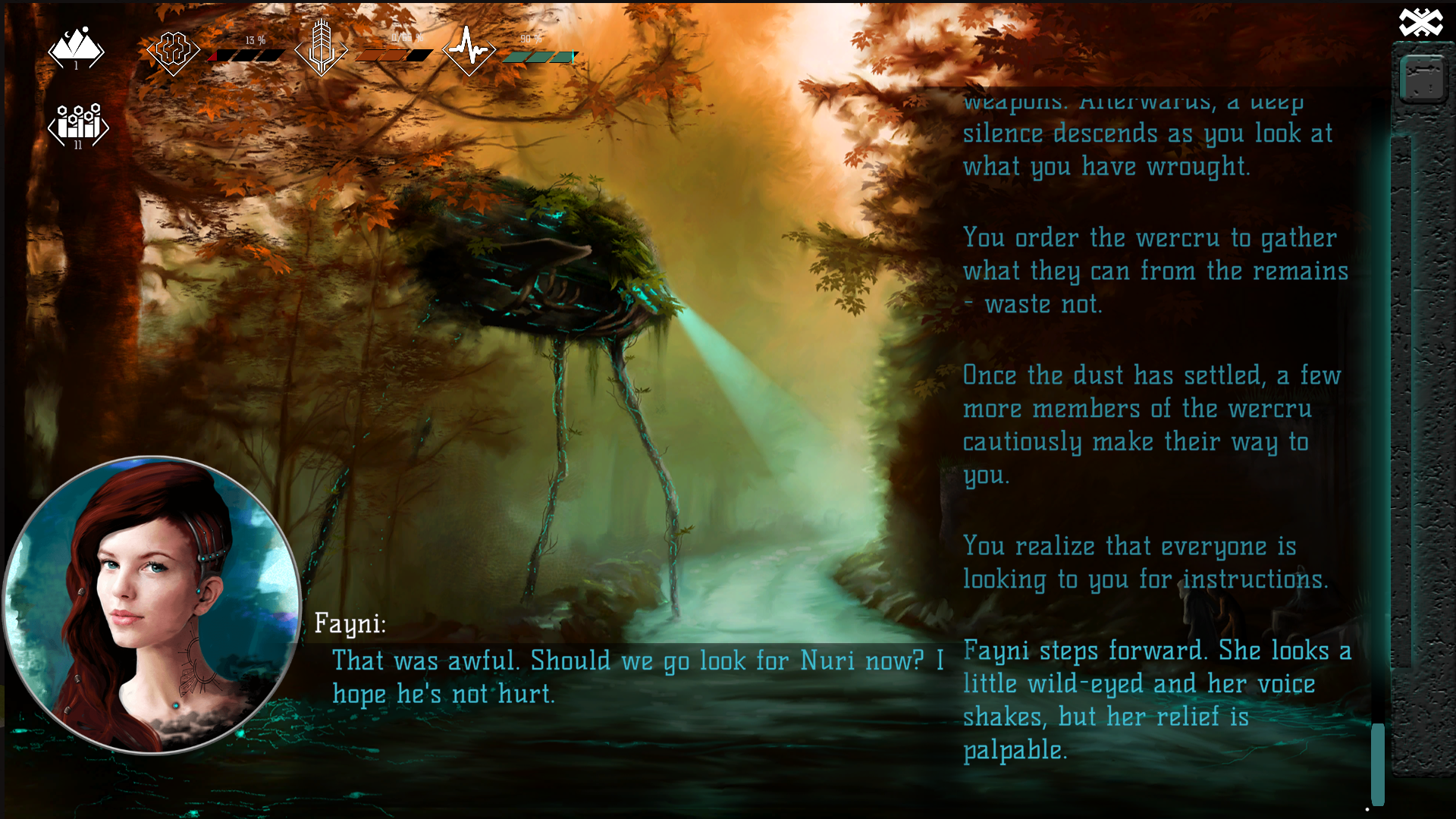}\,
\caption{Screenshots from the game \textit{C.L.A.Y.} (2021). In the story-driven game \textit{C.L.A.Y.} the storyline is sourced from simulations of quantum devices.}
\label{screenshotsclay}
\end{figure*}

\textit{C.L.A.Y.} is a game challenging the notion of using quantum computers for games, and for which a demo was released by MiTale in 2021 \cite{mitale2021clay}. In \textit{C.L.A.Y.}, the dynamics of the storyline rely heavily on the emulation of interference effects on a quantum computer, thereby shaping the interactive storytelling experience (see Figure \ref{screenshotsclay}) \cite{skult2022}. The game's environment design and some visual effects are based on procedural generation using quantum computer-generated randomness, but there are no visual references to either quantum computing or quantum physics in the game itself, nor is there a scientific purpose behind developing this game. Due to its strong dimension of quantum technologies, this game is, to our knowledge, the first-ever commercial quantum computer game.

\subsubsection{The Dimension of Scientific Purposes}
A game might lack visual or otherwise perceivable references to quantum physics, nor run on a quantum computer, yet still have a solid connection to quantum physics through other means, such as by simulating a quantum mechanical phenomenon numerically. Examples of such games have been developed at the \textit{Quantum Game Jam} events, where a special tool called the \textit{Quantum Black Box}, developed for calculating a numerical simulation of the dynamics of a quantum mechanical particle through its time-dependent Schrödinger’s equation, was incorporated into games \cite{qbb,kultima2021qgj,piispanen2023projects, piispanen2024thesis}. From the \textit{Quantum Game Jam} games that incorporated the tool, for games like \textit{Quantum Sheep} (2016), \textit{Quantum Cabaret} (2019), and \textit{Quantum Fruit} (2019) the connection is not necessarily explained, nor are there any clearly visible references to quantum physics in the games themselves. Yet, \textit{Quantum Fruit} worked as a proof-of-concept for the use of the tool in later projects \cite{piispanen2024thesis}. 

The motivation for developing these games has been to create prototypes of citizen science games of the sort of \textit{Quantum Moves 2} and \textit{meQuanics} \cite{quantummoves, devitt2016, piispanen2024cisci}. Therefore, inspired by these, we introduce a third layer that characterises the \textbf{dimension of scientific purposes} (in the field of quantum physical sciences). In the context of quantum physics-related games, the reference to science is, of course, specifically the science of studying quantum physics, quantum technologies, and quantum computing. In addition to the citizen science motivation, other scientific purposes include serious game motivations like education, outreach and testing the capabilities or limitations of a quantum computer, like in the puzzle game \textit{Quantum Awesomeness} from 2018 \cite{wootton2018, wootton2022}.  %(see Figure \ref{screenshotsqc}c )
An educational game like \textit{Quantum Game} by \textit{Quantum Flytrap} has the scientific purpose of educating the player about quantum optics in scientifically accurate ways \cite{migdal2022visualizing}. Therefore, \textit{Quantum Game} also fits the dimension of scientific purposes.\\

Although some quantum physics themed games might not have been successful in their original purpose as serious games, we still want to include them under the definition of quantum games through this dimension. This aspect touches many of the games developed at quantum physics themed game jams and hackathons. Therefore, we define that \textit{if a game is intended as an educational game, a citizen science game, uses a tool designed for such games, or otherwise has a purpose toward a scientific use in the study of quantum physics, it has a dimension of scientific purposes}. Whether these games prove helpful in classrooms or as stand-alone games, requires additional attention from the educator. One of the strongest examples of a game with quantum physics-related scientific purposes is \textit{Quantum Moves 2}, where player solutions can be employed as seeds for numerical optimisation algorithms \cite{jensen2021}. Similarly, an educational game with a focus on quantum literacy\footnote{(as defined in \cite{nita2021})} that had proven to teach the player about quantum physics would have a strong dimension of scientific purposes in addition to having a strong perceivable dimension of quantum physics. 

\subsection{Definition of a Quantum Game}
During the years of developing quantum physics-related games and games on quantum computers, learning about quantum physics-related games made by others, and coming together to dissect and discuss them, the authors found three distinctive dimensions prominent in the games and the motivations behind their development. From our findings and discussion, we distil the definition of \textit{quantum games} into the following statement:

In the context of video games, card and board games, or other games, \textit{quantum games are games that reference the theory of quantum physics, quantum technologies, or quantum computing through perceivable means, connect to quantum physics through a scientific purpose or use quantum technologies.} (See Table \ref{table:definition}) 
\begin{table}[!ht]
\caption{The dimensions of quantum games with descriptions.}
\label{table:definition}
\setlength{\tabcolsep}{6pt} % Default value: 6pt
\renewcommand{\arraystretch}{1.3} % Default value: 1
    \centering
    \begin{tabular}{|l|l|}
    \hline 
        \textbf{DIMENSIONS OF QUANTUM GAMES} \\ \toprule\hline
        \textbf{Perceivable dimension of quantum physics:} \\ 
        The reference to quantum physics in the game is perceivable by interacting 
        with the game \\or with its peripheral material 
        (such as rule books, developers' descriptions, etc.) \\ \hline
        \textbf{Dimension of quantum technologies:} \\
        The game incorporates use of quantum software or quantum 
        devices either during the \\gameplay itself, or during the development of the game \\ \hline 
        \textbf{Dimension of scientific purposes (in the field of quantum physical sciences):} \\
        The game is intended to be an educational game, a citizen science game, 
        uses a tool \\ designed for such games, or otherwise has a purpose 
        towards a scientific use in the study \\of quantum physics \\ \hline
    \end{tabular}
\end{table}

This definition serves both to analyse and understand the multitude of quantum games and to facilitate the development of meaningful quantum games from the perspective of quantum physical sciences. In particular, we note that there exist quantum games that are not science games (i.e., games that serve a scientific purpose) but may instead incorporate purely inspirational aspects intended for entertainment, which are still meaningfully derived from the world of quantum sciences or numerical simulations of quantum physical phenomena. The dimensions presented here are also welcomed to be used in defining and exploring quantum art, quantum music, quantum performing arts, or other creations that reference quantum physics.

\section{Discussion}
Our characterisation of quantum games using the proposed dimensions is intended to lay the groundwork for future research and development of quantum games by providing a broader understanding of how quantum games are part of the ludosphere of games. Future quantum game development may reveal more dimensionality and varying needs for distinction. To give an example, games with a perceivable dimension of quantum physics could be further dissected into games with inspirational and imitative aspects of quantum physics, games where the visuals are derived from numerical simulations of quantum mechanical systems, and games that include clear perceivable educational aspects, if these are important for a specific purpose. Most importantly, we hope that our dimensions of quantum games will offer a tool for incorporating deeper aspects of quantum physics into game development, providing clarity on the different ways a game can be considered a quantum game. \\

The three dimensions of quantum games become particularly interesting when considering the games on the outlines of these dimensions. As an example, the game \textit{Quantum Break} references quantum physics but uses no quantum technologies in gameplay, nor is it intended for educational purposes or citizen science (see Table \ref{table:examples}). It does not fit the early definitions of a quantum game. Still, the core idea of \textit{Quantum Break} and certain visual elements of the game and were designed in collaboration with a quantum physicist \cite{rasanen2016,kamen2016}. We may conclude that \textit{Quantum Break} is a quantum game due to its carefully designed, perceivable references to quantum physics. Certain games, such as \textit{Hopping Mode} by PiCycle from 2014, where the player aids a hovering object through an evolving maze, leave players with only speculative guesses about how the game might relate to quantum physics. The game was developed at the \textit{Quantum Game Jam} of 2014 and has credited quantum physics researchers in the development process, but no other concrete references to quantum physics are found in relation to the game \cite{kultima2021qgj}.

In general, it seems common for a quantum game to feature a perceivable dimension of quantum physics. However, as we have seen, there are games that lack a clear perceivable dimension of quantum physics but include either a dimension of quantum technologies or scientific purposes. Among all the studied quantum games that both served a scientific purpose and used quantum technologies, no example was found of a game lacking a perceivable dimension of quantum physics, completely from the game or its peripheral materials. This is likely due to the fact that quantum technologies are regarded as an interesting novel resource for games and are therefore heavily underlined, whenever used. In many cases, the idea of creating a quantum game is the very reason that these games were created. This causes a natural tendency to prominently feature the quantum aspect of the game. Additionally, using current prototype quantum technology or designing the game for scientific purposes introduces limitations on the game design. If these limitations are noticeable to the player, it will be obvious that there is something unique about the game. By making the quantum aspect perceivable, a context is provided for the players to better understand their experience. This discussion led the authors to converse the design of such a game and play with the idea of what it would mean in practice for a game to have both a dimension of quantum technologies and scientific use, but no perceivable elements referring to quantum physics. The authors invite anyone to approach this challenge.\\

We note that the presented dimensions can also be used to define and explore quantum art, quantum music, quantum performing arts or other creations referencing quantum physics. We also extend this discussion to other scientific fields using the presented dimensions, focusing on perceivable aspects, relations to scientific uses, and potential technologies and simulations central to each field. We propose that games related to biology, astrophysics or cosmology, for example, could be characterised in this manner. We also invite further investigation into additional layers of the dimension of scientific purposes. As for now, most serious quantum games are educational, and only a few have citizen science-related objectives \cite{ahmed2021PHD, piispanen2023projects, piispanen2024cisci, piispanen2024thesis}. Therefore, it does not yet make sense to separate the different types of serious games, such as those for training, citizen science, awareness, education, and outreach, within this dimension when considering quantum games. However, we suspect that the future of quantum games may require a revisit on this dimension. We propose that acknowledging these further layers would rationalise and organise the design objectives and the development of serious quantum games. 

\subsection{The Future of Quantum Games}
In our definition, we did not separate the simulations of quantum technologies (like quantum software) from the usage of quantum technologies (like quantum computers) but rather aligned them along the same dimension. This decision is based on the fact that quantum computers are still in their early stages, and running programs on them results in some degree of error. Quantum software that simulates these devices calculates the same outcomes in the end and often still produces cleaner results in certain types of encoding. The “deeper” end of this dimension is open to debate and welcomes the use of quantum technologies in future quantum games. We expect that, eventually, games and quantum games may merge, and that games might incorporate quantum enhancements in a manner similar to how graphics cards evolved as additional computational resources.

As discussed in Section \ref{s:digi}, the development of early computer games was primarily motivated by outreach purposes. We have observed similar early steps in the development of quantum computer games as they have evolved from implementations of known, existing games on quantum computers to the first prototypes designed for entertainment purposes \cite{wootton2022}. Digital games have evolved enormously since the time of \textit{Spacewar!} and have even driven the development of various gaming hardware and computer components. What might quantum technologies, or quantum computers in particular, offer to games? One such direction could be the vision behind the game \textit{C.L.A.Y.}, which aims to provide unique playing experiences through the use of quantum technologies (see Figure \ref{screenshotsclay}) \cite{mitale2021clay}. If the interaction between the player and a quantum computer could be offered as an immediate, constant feedback loop, the probabilistic nature of quantum phenomena might offer unique feedback on each game run \cite{skult2022}. First patent applications have been filed on methods using quantum computing for team composition modelling in Multiplayer Online Battle Arenas (MOBA) \cite{ebet}. The same companies behind these games also provide online gambling games. The ethics on the use of quantum computers for such purposes should be discussed openly.

A near-term application of quantum computers could be found in the study of procedural content generation (PCG) \cite{wootton2020a,wootton2021,heese2023}. PCG is employed for various purposes across many game genres. This includes classical roguelikes and modern roguelites, 4X strategy games such as the \textit{Civilization} series, and sandbox games such as \textit{Minecraft} or \textit{No Man's Sky}. When designing a content generator, one faces a number of trade-offs, where qualities such as controllability, speed, and robustness of the generator are in partial conflict \cite{togelius2011}. Consequently, content generators for large areas and complex systems are therefore typically limited in ambition and/or expressivity. For example, generating a large open-world environment often makes it computationally infeasible to check for playability constraints such as blocked paths or inaccessible materials. Therefore, PCG designer will typically limit the generator's output range to ensure robustness, i.e., to avoid game-breaking content. In this context, quantum computing might enable the development of content generators that design larger, more systematic content while maintaining playability constraints. This likely requires encoding dependencies between content features in a manner that quantum computers can resolve. Many PCG problems could be posed as constraint satisfaction problems, and solved using e.g. Answer Set Programming, or evolutionary algorithms \cite{smith2011}. 

We acknowledge that our early preprint has already inspired discussion upon this topic \cite{skult2024chronicle}, but much more is to be investigated about the possibilities of using quantum computers as the source of the randomness in PCG. Random numbers play important roles in cryptography, scientific and applied simulations, but also in creating an immersive experience in games. Algorithmic-based, computer-generated randomness is never truly random in the strict sense of the term and can eventually lead to sequences appearing in the outcome, noticeable for a keen eye. Methods where environmental factors have been incorporated into the seed or into the steps of the random number algorithms have been able to provide more than just pseudo-randomness, but quantum computers may offer truly random numbers without extra trickery or biassed outcomes because the outcome of a measurement in quantum physics is based on a fundamentally random phenomena.\\

For future work, a more thorough and systematic analysis of the collection of quantum games and their dimensions should be addressed, in particular for the purposes of education. Although the analysis of the selected games was deemed adequate for the purpose of forming a suitable definition, a closer examination of them may reveal more characteristics of the ludosphere of quantum games. There is also a lot of room for discussion when it comes to games that use well-defined numerical quantum simulations or quantum technologies for the purpose of random number generators. Do these games qualify as quantum games? We argue they do if their usage aligns with any of the presented dimensions of quantum games. Moreover, these specific examples underline the importance of familiarisation and education of the teams developing quantum games. In order to develop more than just a fancy random number generator, the development process requires expertise in both quantum physics and game design \cite{piispanen2023projects, piispanen2024thesis}. 

\section{Conclusion}
In this article, we define \textit{quantum games} through their characteristics. In our model, quantum games are analysed and defined by a set of three \textit{dimensions of quantum games} described as the \textit{perceivable dimension of quantum physics}, the \textit{dimension of quantum technologies}, and the \textit{dimension of scientific purposes} (in the field of quantum physical sciences). The dimensions presented here are also welcomed to be used in defining and exploring quantum art, quantum music, quantum performing arts, or other creations that reference quantum physics.\\

A game exhibits a \textit{perceivable dimension of quantum physics} if the reference to quantum physics in the game is perceivable by interacting with the game or with its peripheral material (such as rule books, developers' descriptions, etc.). The \textit{dimension of quantum technologies} is characterised by the use of quantum software or devices, either during gameplay or in the development of the game. \textit{Quantum computer games}, defined as games playable on quantum computers, would present this dimension, highlighting the novelty of creating games using quantum computers. The \textit{dimension of scientific purposes} (in the field of quantum physical sciences) is defined by the statement that the game is intended to be an educational game, a citizen science game by using a tool designed for such games, or otherwise has a purpose towards a scientific use in the study of quantum physics. The game exhibiting this dimension would thus also fall under our definition for a \textit{science game}, defined in this article as \textit{a game contributing to scientific work, either directly (such as helping to solve research questions) or indirectly (such as building awareness, training, or teaching a scientific topic)}. This definition includes citizen science games, games for outreach and other science-related serious-purpose games, not just educational ones, under the concept of science games.\\

A game may exhibit one or more dimensions of quantum
games and therefore would be called a \textit{quantum game}. Our dimensions provide a tool to evaluate and dissect a game as a quantum game, and offer concrete directions for designing a game with a profound connection to quantum physics. 

With more than 350 quantum physics-related games developed by the end of summer 2024, we note that not all quantum games are science games or playable on quantum computers, but may still have aspects that are “inherently quantum” \cite{quantumgames}. The number of quantum games is steadily growing as quantum computing progresses toward a playable future, and it is anticipated that quantum games will take further forms, similar to the expansion of the general ludosphere of games. %The framework of the presented dimensions are viewed strictly from the view of quantum physics-related sciences, but we want to poin
\backmatter

%\bmhead{Supplementary information}
%If your article has accompanying supplementary file/s please state so here. 

\section*{Declarations}
%Some journals require declarations to be submitted in a standardised format. Please check the Instructions for Authors of the journal to which you are submitting to see if you need to complete this section. If yes, your manuscript must contain the following sections under the heading `Declarations':
\begin{itemize}
\item \textbf{Funding}: LP acknowledges financial support from the Research Council of Finland PROFI funding under the Academy decision number 318937, the Vaisala Foundation, Alfred Kordelin Foundation, and the Finnish Quantum Institute, InstituteQ, during the duration of the presented study. 
JW acknowledges support from the NCCR SPIN, a National Centre of Competence in Research, funded by the Swiss National Science Foundation (grant number 51NF40-180604).
\item \textbf{Acknowledgements}: A special thanks to the group of QWorld interns, who combined the unofficial list of quantum games gathered by LP in 2019, the \textit{Awesome Quantum Games} \cite{awesome} and added valuable new resources to them. We greatly appreciate the input of Dr. Tom Bullock into proofreading our article.
\item \textbf{Conflict of interest/Competing interests} Not applicable
\item \textbf{Ethics approval and consent to participate}: Not applicable
\item \textbf{Consent for publication}: Consent granted by all the authors.
\item \textbf{Data availability}: All the games related to this research are openly listed at \cite{quantumgames}. For viewing the table used for finding the aspects and dimensions available through the corresponding author.
\item \textbf{Materials availability}: The games used for the analysis are mostly available through the reference at \cite{quantumgames}.
\item \textbf{Code availability:} Not applicable

\item \textbf{Author contribution}: LP: Designed the study including the methods for game analysis, collection of the games and structuring the writing of this article. She gathered and characterised the games, gathered the literature review, maintained article integrity, drafted and expanded sections on digital game history and quantum computer games based on JW's work, led discussions on science games and games for serious purposes, suggested the model of dimensions, and contributed to the structure and naming of proposed dimensions. She has been responsible for the overall integrity of the article through its several iterations as the corresponding author.

MP: Contributed to updating the list of games, discussing the games and their dimensions, and carefully revising earlier versions of the article.

AK: Gathered academic literature on game definitions and play, and contributed to discussions on the structure and naming of proposed dimensions.

LJ, MP, JW, JT, AK: Discussed game analysis and interpretation of results, and contributed substantially to the Discussion section by critically examining the overall process of the analysis and the future directions of quantum games.
\end{itemize}
\noindent
%If any of the sections are not relevant to your manuscript, please include the heading and write `Not applicable' for that section. 

%%===================================================%%
%% For presentation purpose, we have included        %%
%% \bigskip command. Please ignore this.             %%
%%===================================================%%
\begin{appendices}

\section{Tables related to classifying quantum physics-related games according to the dimensions of the quantum games}
\label{sec:A1}
\begin{table}[!ht]
\caption{A comprehensive table of quantum games referenced in Section \ref{sec:dimensions}, categorised by the dimensions of quantum games, in the order they were mentioned. The table includes 16 example games selected from the 250 games examined between 2019 and 2022, characterised according to the three dimensions of quantum games. For each listed game, a Yes/No indicator is provided to show whether the game exhibits the qualities described by the three dimensions: the perceivable dimension of quantum physics (“Perceivable”), the dimension of quantum technologies (“Q Tech”), and the dimension of scientific purposes (“Sci purpose”). In the “Perceivable” dimension, “simulation” is noted for games that numerically simulate perceivable quantum physical phenomena on a classical computer. In the final column, the purposes are noted: “Edu” for educational purposes, “Citizen Sci” for citizen science and “Benchm” that the game has been designed for benchmarking quantum computers. For the game \textit{Hopping Mode}, no information was available about the original purpose of the game, although it was developed during a \textit{Quantum Game Jam} event.
 }
	\label{table:exampleslong}
 \setlength{\tabcolsep}{6pt} % Default value: 6pt
\renewcommand{\arraystretch}{1.3} % Default value: 1
	\centering
\begin{tabular}{|llll|}
\hline
Game                     & Perceivable  & Q Tech & Sci purpose  \\
\toprule
\hline
Quantum Moves 2          & Yes (simulation)  & No	& Yes (Citizen Sci)   \\
Hamsterwave              & Yes (simulation)  & No	& Yes (Citizen Sci)   \\
Quantum TiqTaqToe        & Yes (simulation)  & No	& Yes (Edu)   \\
Quantum Game  	         & Yes (simulation)  & No 	& Yes (Edu)   \\
Quantum Entanglement     & Yes   & No	& No   \\
Quantum Labyrinth        & Yes   & No	& No  \\
Escape from the Quantum Computer & Yes   & No	& No  \\
Cat/Box/Scissors	     & Yes	 & Yes  & No \\
C.L.A.Y.		         & No    & Yes  & No \\
Quantum Sheep 	         & No (simulation)   & No   & Yes (Citizen Sci) \\
Quantum Cabaret 	         & Yes (simulation)    & No   & Yes (Citizen Sci) \\
Quantum Fruit 		     & No (simulation)   & No  & Yes (Citizen Sci) \\
meQuanics		         & Yes (simulation)	 & No	& Yes (Citizen Sci)  \\
Quantum Awesomeness	     & Yes	 & Yes	& Yes (Benchm)  \\
Quantum Break	         & Yes   & No   & No  \\
Hopping Mode	         & No   & No   & ?   \\
 \hline  
\end{tabular}
\end{table}

\begin{table}[!ht]
	\caption{The questions used to analyse the list of quantum games as described in Section \ref{sec:methods}. For some 1/0 type answers, commenting was used for clarification.}
	\label{table:finalquestions}
 \setlength{\tabcolsep}{6pt} % Default value: 6pt
\renewcommand{\arraystretch}{1.3} % Default value: 1
	\centering
\begin{tabular}{|ll|}
\hline 
\textbf{Purpose} (intent for a serious use)& type \\ 
\hline  
(has the game been developed for) Benchmarking (q. computers) & 1/0\\
(has the game been developed for) Citizen Science & 1/0\\
Separate educational tutorial/instruction & 1/0\\
Educational aspects in the game/tutorial & 1/0\\
Concepts you learn about and are explained correctly & text\\
Concepts you witness being used but are not explained well & text \\
Clear misconceptions & 1/0\\
\hline

\hline 
\textbf{Perceivable aspects} & type \\ 
\hline  
(Does the) Backstory or storyline reference (QP) & 1/0\\
Only the related material connect to QP & 1/0\\
(Is there a) Quantum physicists in the team or recognised being involved & 1/0\\
Are there literal to be seen representation of quantum; particles, quantum computers, & 1/0\\ gates, refrigerators, a lab, a Schrödinger cat, etc &\\
Effects to be seen, actions that are quantum or “quantum” & 1/0\\
Superposition presented as “something in two or more places in one time” & 1/0\\
Entanglement presented as “two objects move simultaneously” or “to same/opposite directions”  & 1/0 \\
(Are there) Visuals generated by the use of numerical simulations of QP phenomena& 1/0\\
\hline

\hline 
\textbf{Q Technologies (in the development and/or in gameplay)} & type \\ 
\hline  
Music quantum computer generated & 1/0\\
Visuals quantum computer generated & 1/0\\
Quantum procedural generation & 1/0\\ 
Runs on a quantum computer (at any point) & 1/0\\
Quantum computer has been used for the development & 1/0\\
Q tech / numerics are used only like a RNG & 1/0 \\
\hline

\hline 
\textbf{Simulation (in the development and/or in gameplay)} & type \\ 
\hline  
Quantum computing software used ( Qiskit etc ) & 1/0\\
A numerical simulation has been used (QBB etc) & 1/0\\
\hline
\end{tabular}
\end{table}
%An appendix contains supplementary information that is not an essential part of the text itself but which may be helpful in providing a more comprehensive understanding of the research problem or it is information that is too cumbersome to be included in the body of the article.

%%=============================================%%
%% For submissions to Nature Portfolio Journals %%
%% please use the heading ``Extended Data''.   %%
%%=============================================%%

%%=============================================================%%
%% Sample for another appendix section			       %%
%%=============================================================%%

%% \section{Example of another appendix section}\label{secA2}%
%% Appendices may be used for helpful, supporting or essential material that would otherwise 
%% clutter, break up or be distracting to the text. Appendices can consist of sections, figures, 
%% tables and equations etc.
\pagebreak
\end{appendices}

%%===========================================================================================%%
%% If you are submitting to one of the Nature Portfolio journals, using the eJP submission   %%
%% system, please include the references within the manuscript file itself. You may do this  %%
%% by copying the reference list from your .bbl file, paste it into the main manuscript .tex %%
%% file, and delete the associated \verb+\bibliography+ commands.                            %%
%%===========================================================================================%%

\bibliography{references}% common bib file
%% if required, the content of .bbl file can be included here once bbl is generated
%%\input sn-article.bbl

\end{document}